\documentclass[graybox]{svmult}

\usepackage{mathptmx}       
\usepackage{makeidx}         
\usepackage{graphicx}        
\usepackage{cite}            
\usepackage[bottom]{footmisc}
\usepackage{amsbsy,amssymb,amsmath,bm}

\makeindex             

\begin{document}

\title*{Orbital fluctuations in the $R$VO$_3$ perovskites}

\author{Andrzej M. Ole\'s and Peter Horsch}

\institute{Andrzej M. Ole\'s \at
           M. Smoluchowski Institute of Physics, Jagellonian University,
           Reymonta 4, 30059 Krak\'ow, Poland \\
           Max-Planck-Institut f\"ur Festk\"orperforschung,
           Heisenbergstrasse 1, 70569 Stuttgart, Germany \\
           \email{A.M.Oles@fkf.mpg.de}
      \and Peter Horsch \at
           Max-Planck-Institut f\"ur Festk\"orperforschung,
           Heisenbergstrasse 1, 70569 Stuttgart, Germany \\
           \email{P.Horsch@fkf.mpg.de}}

\maketitle

\abstract{
The properties of Mott insulators with orbital degrees of freedom
are described by spin-orbital superexchange models, which provide
a theoretical framework for understanding their magnetic and optical 
properties. We introduce such a model derived for
$(xy)^1(yz/zx)^1$ configuration of V$^{3+}$ ions in the $R$VO$_3$
perovskites, $R$=Lu,Yb,$\cdots$,La, and demonstrate that $\{yz,zx\}$ 
orbital fluctuations along the $c$ axis are responsible for the huge 
magnetic and optical anisotropies observed in the almost perfectly 
cubic compound LaVO$_3$. We argue that the GdFeO$_3$ distortion and the
large difference in entropy of $C$-AF and $G$-AF phases is responsible 
for the second magnetic transition observed at $T_{N2}$ in YVO$_3$. 
Next we address the variation of orbital and magnetic transition 
temperature, $T_{\rm OO}$ and $T_{N1}$, in the $R$VO$_3$ perovskites, 
after extending the spin-orbital model by the crystal-field and the 
orbital interactions which arise from the GdFeO$_3$ and Jahn-Teller 
distortions of the VO$_6$ octahedra. We further find that the 
orthorhombic distortion which increases from LaVO$_3$ to LuVO$_3$ plays 
a crucial role by controlling the orbital fluctuations, and via the 
modified orbital correlations influences the onset of both magnetic 
and orbital order.
}

\section{Orbital degrees of freedom in strongly correlated systems}
\label{sec:orbi}

Orbital degrees of freedom play a key role for many intriguing 
phenomena in strongly correlated transition metal oxides, such as the
colossal magnetoresistance in the manganites or the effective reduction
of dimensionality in KCuF$_3$ \cite{Tok00}. Before addressing complex
phenomena in doped Mott insulators, it is necessary to describe first 
the undoped materials, such as LaMnO$_3$ or LaVO$_3$. These two systems 
are canonical examples of correlated insulators with coexisting 
magnetic and orbital order \cite{Tok00}. In both cases large local 
Coulomb interaction $U$ suppresses charge fluctuations, leading to 
low-energy effective Hamiltonians with superexchange interactions which 
stabilize antiferromagnetic (AF) spin order at low temperature 
\cite{Fei99,Kha01}.
However, the AF order is different in both cases: ferromagnetic (FM)
planes are coupled by AF interactions in the $A$-type AF phase of
LaMnO$_3$, while FM chains along the $c$ cubic axis are coupled by
AF interactions in the $ab$ planes in the $C$-type AF ($C$-AF)
phase of LaVO$_3$. The superexchange Hamiltonians which describe both 
systems are just examples for the spin-orbital physics \cite{Ole05},
where orbital (pseudospin) operators contribute explicitly to the
structure of the superexchange interactions --- their actual form
depends on the number of $3d$ electrons (holes) at transition metal
ions which determines the value of spin $S$, and on the type of
active orbital degrees of freedom, $e_g$ or $t_{2g}$.
In simple terms, the magnetic structure is determined by the pattern 
of occupied and empty orbitals, and the associated rules are known as 
Goodenough-Kanamori rules (GKR). The central focus of this overview 
are $t_{2g}$ orbital degenerate systems, where quantum fluctuations
of orbitals play a central role for the electronic properties 
\cite{Kha01,Miy05,Yan07} and modify the predictions of the GKR.

In the last two decades several new concepts were developed in the
field of orbital physics \cite{Tok00}. The best known spin-orbital
superexchange Hamiltonian is the Kugel-Khomskii model \cite{Kug82}, 
which describes the $e_g$ orbital $\{x^2-y^2,3z^2-r^2\}$ degrees of 
freedom coupled to $S=1/2$ spins at Cu$^{2+}$ ($d^9$) ions in KCuF$_3$. 
The spins interact by either FM and AF exchange interactions, 
depending on the type of occupied and empty orbitals on two neighboring 
ions. It has been found that enhanced quantum fluctuations due to 
orbital degrees of freedom, which contribute to joint spin-orbital 
dynamics, may destabilize long-range magnetic order near the quantum 
critical point of the Kugel-Khomskii model \cite{Fei97}. The orbital 
part of the superexchange is thereby intrinsically frustrated even on 
geometrically non-frustrated lattices, as in the perovskite lattice 
\cite{Fei97,vdB04}, which is a second important concept in the field of
orbital physics. Finally, although spin and orbital operators commute,
there are situations where joint spin-orbital dynamics plays a crucial
role, and spin and orbital operators cannot be separated from each
other. This situation is called spin-orbital entanglement \cite{Ole06},
and its best example are the entangled SU(4) singlets in the
one-dimensional (1D) SU(4) model \cite{Fri99}.
There is no doubt that these recent developments in the orbital
physics provide many challenges both for the experimental studies and
for the theoretical understanding of the experimental consequences of
the spin-orbital superexchange.

Let us consider first the orbital part of the superexchange. Its
intrinsic frustration results from the directional nature of orbital
pseudospin interactions \cite{Fei97,vdB04} --- they imply that the pair
of orbitals which would minimize the energy depends on the direction of
a bond $\langle ij\rangle$ in a cubic (peovskite) lattice. In case of 
$e_g$ orbitals the superexchange interactions are Ising-like as only one 
orbital flavor allows for electron
hopping $t$ and the electron exchange process does not occur. They favor 
a pair of orthogonal orbitals on both sites of the considered bond 
\cite{vdB99}, for instance $|z\rangle\sim (3z^2-r^2)/\sqrt{6}$ and 
$|x\rangle\sim x^2-y^2$ orbital for a bond along the $c$ axis. When the 
two above orbital states are represented as components of $\tau=1/2$ 
pseudospin, this configuration gives the energy of $-\frac{1}{4}J$, 
where $J$ is the superexchange constant. Unlike in the 1D model 
\cite{Dag04}, such an optimal orbital configuration cannot be 
realized simultaneously on all the bonds in a two-dimensional (2D) or 
three-dimensional (3D) system. Thus, in contrast to spin systems, the 
tendency towards orbital disordered state (orbital liquid) is 
{\it enhanced\/} with increasing system dimension \cite{Fei05,Kha05}.

The essence of orbital frustration is captured by the 2D compass model,
originally developed as a model for Mott insulators \cite{Kug82}.
Intersite interactions in the compass model are descibed by products
$\tau^{\alpha}_i\tau^{\alpha}_j$ of single pseudospin components,
\begin{equation}
\tau^x_i = \frac{1}{2}\sigma^{x}_i , \hskip 1cm
\tau^y_i = \frac{1}{2}\sigma^{y}_i , \hskip 1cm
\tau^z_i = \frac{1}{2}\sigma^{z}_i .
\label{t2g}
\end{equation}
for a bond $\langle ij\rangle\parallel\gamma$, where $\alpha=x,y,z$,
rather than by a pseudospin scalar product
${\vec \tau}_i\cdot{\vec \tau}_j$. For instance, in the 2D case of a
single $ab$ plane, the compass model \cite{Kho03},
\begin{equation}
H_{2D} =
 J_x\sum_{\langle ij\rangle\parallel a}\tau^x_i\tau^x_j
+J_z\sum_{\langle ij\rangle\parallel b}\tau^z_i\tau^z_j\;,
\label{compa}
\end{equation}
describes the competition between $\tau^x_i\tau^x_j$ and
$\tau^z_i\tau^z_j$ interactions for the bonds along $a$ and $b$ axis,
respectively. This competition of pseudospin interactions along
different directions results in intersite correlations similar to those
in the anisotropic XY model, and generates a quantum critical point at
$J_x=J_z$, with high degeneracy of the ground state \cite{Mil05}.
So, despite certain similarities of the compass model to ordinary
models used in quantum magnetism, an ordered phase with finite 
magnetization is absent. It is interesting to note that a similar
quantum phase transition exists also in the 1D chain compass model
\cite{Brz07} ($N'=N/2$ is the number of unit cells):
\begin{equation}
H_{1D}=\sum_{i=1}^{N'}
\left\{ J_x\tau_{2i-1}^x \tau_{2i}^x +
        J_z\tau_{2i}^z   \tau_{2i+1}^z \right\}\,.
\label{H1}
\end{equation}
Recently this 1D compass model was solved exactly in the whole range
of $\{J_x,J_z\}$ parameters \cite{Brz07} by mapping to the exactly
solvable (quantum) Ising model in transverse field. It provides a 
beautiful example of a first order quantum phase transition between 
two phases with large $\langle\tau_{2i-1}^x \tau_{2i}^x\rangle$ or 
$\langle\tau_{2i}^z \tau_{2i+1}^z\rangle$ correlations, and 
discontinuous changes of intersite correlation functions.

In realistic spin-orbital superexchange models transitions between
different ordered or disordered orbital states are accompanied by
magnetic transitions. This field is very rich, and several problems
remain unsolved as simple mean-field (MF) approaches do not suffice in
general, even for the systems with perovskite lattices \cite{Ole05}. 
In this chapter we shall address the physical properties of the 
$R$VO$_3$ perovskites ($R$=Lu,Yb,$\cdots$,La), where not only the above 
intrinsic frustration of the orbital superexchange, but also the 
structure of the spin-orbital superexchange arising from multiplet 
splittings due to Hund's exchange plays a role and determines the 
observed physical properties at finite temperature. Moreover, we shall 
see that the coupling of the orbitals to the lattice, i.e., via 
Jahn-Teller (JT) coupling, GdFeO$_3$-like and orthorhombic distortion,
are important control parameters. First we analyze the structure of 
the spin-orbital superexchange in section 2 and show its consequences 
for the magnetic and optical properties of strongly correlated
transition metal compounds. Here we also address the entanglement of
spin and orbital variables which is ignored in the MF decoupling, 
and we point out that it fails in certain situations.

\begin{figure}[t!]
\sidecaption[t]
\includegraphics[width=6.8cm]{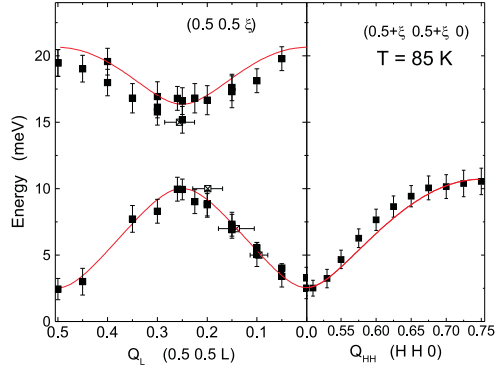}
\caption{Magnon dispersion relation obtained by neutron scattering
for the $C$-AF phase of YVO$_3$ at $T=85$ K. The lines are
interpolation between the experimental points (squares with error
bars) along two high symmetry directions in the Brillouin zone.
Image courtesy of Clemens Ulrich.}
\label{fig:yvo}
\end{figure}

The coupling between the orbital and spin variables is capable of
generating qualitatively new phenomena which do not occur in the
absence of orbital interactions, such as anisotropic magnetic
interactions, and novel quantum phenomena at finite temperature, 
discusseed on the example of LaVO$_3$ in section 3. One of such novel
and puzzling phenomena is the magnetic phase transition between two
different types of AF order observed in YVO$_3$ --- this compound has
$G$-type AF order (staggered in all three directions, called below
$G$-AF phase) at low temperature $T<T_{N2}$, while the magnetic order
changes in the first order magnetic transition at $T_{N2}=77$ K to
$C$-AF phase which remains stable up to $T_{N1}\simeq 116$ K. The
latter $C$-AF phase has rather exotic magnetic properties, and the
magnon spectra show dimerization of the FM interactions along the $c$
axis \cite{Ulr03}, see figure \ref{fig:yvo}. In fact, the $G$-AF phase
occurs in systems with large GdFeO$_3$-like distortion \cite{Miy03}. In
Ref. \cite{Kha01} an orbital interaction favoring $C$-type alternating 
orbital ($C$-AO) order was invoked to explain the $G$-AF phase. We also 
address this problem in section 3 and present arguments that at 
higher $T>T_{N2}$ $C$-AF phase reappears is due to its higher entropy
\cite{Kha01}.

In section 4 we address the experimental phase diagram of the $R$VO$_3$
perovskites. It is quite different from the (also puzzling) phase
diagram of the $R$MnO$_3$ perovskites, where the orbital order (OO)
appears first at $T_{\rm OO}$ upon lowering the temperature, and spin
order follows \cite{Goo06} at the N\'eel temperature,
$T_{N1}\ll T_{\rm OO}$. In contrast, in the $R$VO$_3$ perovskites the
two transitions appear at similar temperature \cite{Miy06}. For 
instance, in LaVO$_3$ they occur even almost simultaneously i.e.,
$T_{N1}\simeq T_{\rm OO}$. However, they become separated from each
other in the $R$VO$_3$ systems with smaller ionic radii of $R$ ions ---
whereas $T_{N1}$ gets reduced for decreasing $r_R$, $T_{\rm OO}$
exhibits a {\it nonmonotonic\/} dependence on $r_R$ \cite{Miy03}.
A short summary is presented in section 5. We also point out a few
unsolved problems of currect interest in the field of orbital physics.

\section{Spin-orbital superexchange and entanglement }
\label{sec:som}

Spin-orbital models derived for real systems are rather complex and the 
orbital part of the superexchange is typically much more complicated 
than the compass model (\ref{compa}) of section 1 \cite{Ole05}. 
The main feature of these
superexchange models is that not only the orbital interactions are
directional and frustrated, but also spin correlations may influence
orbital interactions and {\it vice versa\/}. This is best seen in the
Kugel-Khomskii ($d^9$) model \cite{Kug82}, where the $G$-AF and $A$-AF
order compete with each other, and the long-range order is destabilized
by quantum fluctuations in the vicinity of the quantum critical point
$(J_H,E_z)=(0,0)$ \cite{Fei98}. Here $J_H$ is the local exchange (see
below), and $E_z$ is the splitting of two $e_g$ orbitals. Although this
model is a possible realization of disordered spin-orbital liquid, its
phase diagram remains unexplored beyond the MF approach and simple
valence-bond wave functions of Ref. \cite{Fei97} --- it remains one of
the challenges in this field.

In this chapter we consider the superexchange derived for an (idealized)
perovskite structure of $R$VO$_3$, with V$^{3+}$ ions occupying the
cubic lattice. The kinetic energy is given by:
\begin{equation}
\label{hkin}
H_{t}=-t\sum_{\langle ij\rangle{\parallel}\gamma}\;
         \sum_{\alpha(\gamma),\sigma}
\left(d^{\dagger}_{i\alpha\sigma}d^{}_{j\alpha\sigma}+
      d^{\dagger}_{j\alpha\sigma}d^{}_{i\alpha\sigma}\right),
\end{equation}
where $d^{\dagger}_{i\alpha\sigma}$ is electron creation operator for an
electron with spin $\sigma=\uparrow,\downarrow$ in orbital $\alpha$ at
site $i$.
The summation runs over the bonds $\langle ij\rangle{\parallel}\gamma$
along three cubic axes, $\gamma=a,b,c$, with the hopping elements $t$
between active $t_{2g}$ orbitals. They originate from two subsequent
hopping processes via the intermediate $2p_{\pi}$ oxygen orbital along 
each V--O--V bond. Its value can in principle be derived from
the charge-transfer model \cite{Zaa93}, and one expects
$t=t_{pd}^2/\Delta\sim 0.2$ eV \cite{Kha01}. Only two out of three
$t_{2g}$ orbitals, labelled by
$\alpha(\gamma)$, are active along each bond $\langle ij\rangle$ and
contribute to the kinetic energy (\ref{hkin}), while the third orbital
lies in the plane perpendicular to the $\gamma$ axis and the hopping
via the intermediate oxygen $2p_{\pi}$ oxygen is forbidden by symmetry.
This motivates a convenient notation used below,
\begin{equation}
\label{abc}
|a\rangle\equiv |yz\rangle, \hskip .7cm
|b\rangle\equiv |xz\rangle, \hskip .7cm
|c\rangle\equiv |xy\rangle,
\end{equation}
where the orbital inactive along a cubic direction $\gamma$ is
labelled by its index as $|\gamma\rangle$.

The superexchange model for the $R$VO$_3$ perovskites arises from
virtual charge excitations between V$^{3+}$ ions in the high-spin
$S=1$ state. The number of $d$ electrons is 2 at each V$^{3+}$ ion
($d^2$ configuration), and the superexchange is derived from all
possible virtual $d_i^2d_j^2\rightleftharpoons d_i^3d_j^1$ excitation
processes (for more details see Ref. \cite{Ole07}). It is parametrized
by the superexchange constant $J$ and Hund's parameter $\eta$,
\begin{equation}
\label{sex}
J=\frac{4t^2}{U}\,, \hskip 2cm \eta=\frac{J_H}{U}\,,
\end{equation}
where $U$ is the intraorbital Coulomb interaction and $J_H$ is Hund's
exchange between $t_{2g}$ electrons. Here we use the usual convention 
and write the local Coulomb interactions between $3d$ electrons at 
V$^{3+}$ ions by limiting ourselves to intraorbital and two-orbital 
interaction elements \cite{Ole05}:
\begin{eqnarray}
\label{Hee}
H_{\rm int}&=&
   U\sum_{i\alpha}n_{i\alpha  \uparrow}n_{i\alpha\downarrow}
 +\Big(U-\frac{5}{2}J_H\Big)\sum_{i,\alpha<\beta}n_{i\alpha}n_{i\beta}
-2J_H\sum_{i,\alpha<\beta}\textbf{S}_{i\alpha}\cdot\textbf{S}_{i\beta}
\nonumber \\
&+& J_H\sum_{i,\alpha<\beta}
\left( d^{\dagger}_{i\alpha\uparrow}d^{\dagger}_{i\alpha\downarrow}
      d^{       }_{i\beta\downarrow}d^{       }_{i\beta\uparrow}
     +d^{\dagger}_{i\beta\uparrow}d^{\dagger}_{i\beta\downarrow}
      d^{    }_{i\alpha\downarrow}d^{       }_{i\alpha\uparrow}\right).
\end{eqnarray}
When only one type of orbitals is party occupied (as in the present
case of the $R$VO$_3$ perovskites or in KCuF$_3$), the two parameters
$\{U,J_H\}$ are sufficient to describe these
interactions in Eq. (\ref{Hee}):
($i$) the intraorbital Coulomb element $U$ and
($ii$) the interorbital (Hund's) exchange element $J_H$, where
$\{A,B,C\}$ are the Racah parameters \cite{Gri71}.
In such cases the above expression is exact; in other cases when both
$e_g$ and $t_{2g}$ electrons contribute to charge excitations
(as for instance in the $R$MnO$_3$ perovskites), Eq. (\ref{Hee})
is only an approximation --- the anisotropy on the interorbital
interaction elements has to be then included to reproduce accurately
the multiplet spectra of the transition metal ions \cite{Gri71}.
The intraorbital interaction is $U=A+4B+3C$, while $J_H$ depends on
orbital type --- for $t_{2g}$ electrons one finds \cite{Ole05,Gri71}
$J_H=3B+C$.

The perturbative treatment of intersite charge excitations
$d_i^2d_j^2\rightleftharpoons d_i^3d_j^1$ in the regime of $t\ll U$
leads for the $R$VO$_3$ perovskites (and in each similar case 
\cite{Ole05}) to the spin-orbital superexchange model:
\begin{equation}
\label{HJ}
{\cal H}_J=\sum_{\langle ij\rangle\parallel\gamma}H^{(\gamma)}(ij)
=J\sum_{\langle ij\rangle\parallel\gamma}\left\{
\left({\vec S}_i\cdot {\vec S}_j+S^2\right)
      {\hat J}_{ij}^{(\gamma)} +
      {\hat K}_{ij}^{(\gamma)}      \right\}.
\end{equation}
The spin interactions $\propto {\vec S}_i\cdot {\vec S}_j$ obey the
SU(2) symmetry. In contrast, the orbital interaction operators
${\hat J}_{ij}^{(\gamma)}$ and ${\hat K}_{ij}^{(\gamma)}$ involve
directional (here $t_{2g}$) orbitals on each individual bond
$\langle ij\rangle\parallel\gamma$, so they have
a lower (cubic) symmetry. The above form of the spin-orbital
interactions is general and the spin value $S$ depend on the electronic
configuration $d^n$ of the involved transition metal ions (here $n=2$
and $S=1$). For convenience, we introduced also a constant $S^2$ in the
spin part, so for the classical N\'eel order the first term
$\propto {\hat J}_{ij}^{(\gamma)}$ vanishes.

In the $R$VO$_3$ perovskites one finds the orbital operators
\cite{Ole07}:
\begin{eqnarray}
\label{orbj}
{\hat J}_{ij}^{(\gamma)}&=&
\frac{1}{2}\left\{(1+2\eta r_1)
\left({\vec\tau}_i\cdot {\vec\tau}_j
     +\frac{1}{4}n_i^{}n_j^{}\right)\right.          \nonumber \\
&-&\! \left.\eta r_3
    \left({\vec \tau}_i\times{\vec \tau}_j+\frac{1}{4}n_i^{}n_j^{}\right)
-\frac{1}{2}\eta r_1(n_i+n_j)\right\}^{(\gamma)},             \\
\label{orbk}
{\hat K}_{ij}^{(\gamma)}&=&\!
\left\{\eta r_1
\left({\vec\tau}_i\cdot {\vec\tau}_j+\frac{1}{4}n_i^{}n_j^{}\right)
 +\eta r_3\left({\vec\tau}_i\times {\vec\tau}_j
             +\frac{1}{4}n_i^{}n_j^{}\right)\right.  \nonumber \\
&-&\left.
   \frac{1}{4}(1+\eta r_1)(n_i+n_j)\right\}^{(\gamma)},
\end{eqnarray}
where the scalar product $({\vec\tau}_i\cdot {\vec\tau}_j)^{(\gamma)}$
and the cross-product,
\begin{equation}
\label{exo1}
\left({\vec\tau}_i\times {\vec\tau}_j\right)^{(\gamma)}=
\frac{1}{2}\left(\tau_i^+\tau_j^++\tau_i^-\tau_j^-\right)
+\tau_i^z\tau_j^z\,,
\end{equation}
involve orbital (pseudospin) operators corresponding to two active 
$t_{2g}$ orbitals along the $\gamma$ axis, with 
${\vec\tau}_i=\{\tau_i^+,\tau_i^-\tau_i^z\}$, 
and 
\begin{equation}
\label{tauz}
\tau_i^z=\textstyle{\frac{1}{2}}(n_{i,yz}-n_{i,zx})\,.
\end{equation}
They follow from the structure of local Coulomb interaction (\ref{Hee}).
The latter term (\ref{exo1}) leads to the nonconservation of total 
pseudospin quantum number. Density operators
$n_i^{(\gamma)}$ in Eqs. (\ref{orbj}) and (\ref{orbk}) stand for the
number of $d$ electrons in active orbitals for the considered bond
$\langle ij\rangle$, e.g. $n_i^{(c)}=n_{ia}+n_{ib}$. The coefficients,
\begin{equation}
\label{rr}
r_1=\frac{1}{1-3\eta}\,, \hskip 1.5cm r_3=\frac{1}{1+2\eta}\,,
\end{equation}
follow from the energies of $d_i^2d_j^2\rightleftharpoons d_i^3d_j^1$
excitations in the units of $U$:
($i$) $r_1$ represents the high-spin $^4A_2$ excitation of energy
$(U-3J_H)$, while the low-spin excitations are given by
($ii$) $r_2=1$ originating from the low-spin $^2T_1$ and $^2E$
excitations of energy $U$, and
($iii$) $r_3$ represents the low-spin $^2T_2$ states of energy
$(U+2J_H)$.

Magnetic order observed in Mott insulators is usually understood in
terms of the GKR which are based on the MF 
picture and ignore entangled quantum states. These rules state that the
pattern of occupied orbitals determines the spin structure. For example,
for $180^{\circ}$ bonds (e.g. Mn--O--Mn bonds in LaMnO$_3$) there are 
two key rules: 
($i$) if two partially occupied $3d$ orbitals point towards each other,
the interaction is AF, however,
($ii$) if an occupied orbital on one site has a large overlap with an 
empty orbital on the other site of a bond $\langle ij\rangle$,
the interaction is weak and FM due to finite Hund's exchange.
This means that spin order and orbital order are complementary ---
ferro-like (uniform) orbital (FO) order supports AF spin order, 
while AO order supports FM spin order. Indeed, these celebrated rules 
are well followed in LaMnO$_3$ \cite{Wei04} and in KCuF$_3$
\cite{Ole00}, where strong JT effect stabilizes the orbital order and 
suppresses the orbital fluctuations. The AO order is here robust in the 
FM $ab$ planes, while the orbitals obey the FO order along the $c$ axis, 
supporting the AF coupling and leading to the $A$-AF phase for both 
systems. In such cases the GKR directly apply.
Therefore, one may disentangle the spin and orbital operators, and it 
has been shown that this procedure is sufficient to explain both the
magnetic \cite{Fei99} and optical \cite{Ole05} properties of LaMnO$_3$.

As another prominent example of the Goodenough-Kanamori complementarity
we would like to mention the AF phases realized in YVO$_3$ \cite{Ulr03},
which are the subject of intense research in recent years. 
{\it A priori,\/} the orbital interactions between V$^{3+}$ ions in
$d^2$ configuration obey the cubic symmetry, if the $t_{2g}$ orbitals
are randomly occupied. However, the symmetry breaking at the structural
transition where the symmetry is reduced from cubic to orthorhombic, 
which persists in the magnetic phases, suggests that the electronic 
configuration is different. Indeed, the GdFeO$_3$ distortions in the
$R$VO$_3$ structure break the symmetry in the orbital space, and both
the electronic structure calculations \cite{And07} and the analysis
using the point charge model \cite{Hor08} indicate that the electronic
configuration $(xy)^1(yz,zx)^1$ is induced at every site, i.e.,
\begin{equation}
\label{nn} n_{ic}=1\,, \hskip 1.5cm n_{ia}+n_{ib}=1\,.
\end{equation}
The partly filled $\{a,b\}$ orbitals are both active along the $c$ axis,
and may lead either to FO or to AO order. Indeed, depending on this 
orbital pattern, the magnetic correlations are there either AF or FM,
explaining the origin of the two observed types of AF order:
($i$) the $C$-AF phase, and
($ii$) the $G$-AF phase.
However, the situation is more subtle as both orbitals in the orbital 
doublet $\{|yz\rangle,|xz\rangle\}\equiv\{|a\rangle,|b\rangle\}$ at each 
site $i$ are active on the bonds along the $c$ axis. This demonstrates
an important difference between the $e_g$ (with one electron or one
hole in active $e_g$ orbital at each site \cite{Dag04}) and a $t_{2g}$ 
system, such as $R$VO$_3$ perovskite vanadates, where electrons 
occupying two active $t_{2g}$ orbitals may fluctuate and form an 
{\it orbital singlet\/} \cite{Kha01}. The cubic symmetry is thus broken 
as both orbital flavors are active only along the $c$ axis, and 
the bonds in the $ab$ planes and along the $c$ axis are nonequivalent. 
Consequently, superexchange orbital operators (\ref{orbj}) and 
(\ref{orbk}) take different forms along these two distinct directions,
\begin{eqnarray}
\label{orbjc}
{\hat J}_{ij}^{(c)}&=&
\frac{1}{2}\left\{(1+2\eta r_1)
\left({\vec\tau}_i\cdot {\vec\tau}_j +\frac{1}{4}\right)
-\eta r_3\left({\vec\tau}_i\times{\vec\tau}_j+\frac{1}{4}\right)
-\eta r_1\right\}\,,             \\
\label{orbkc}
{\hat K}_{ij}^{(c)}&=&
\left\{\eta r_1
\left({\vec\tau}_i\cdot {\vec\tau}_j+\frac{1}{4}\right)
 +\eta r_3\left({\vec\tau}_i\times {\vec\tau}_j
             +\frac{1}{4}\right)
-\frac{1}{2}(1+\eta r_1)\right\}\,,  \\
\label{orbja}
{\hat J}_{ij}^{(a)}&=&
\frac{1}{4}\left\{(1-\eta r_3)(1+n_{ib}n_{jb})
-r_1(n_{ib}-n_{jb})^2\right\}\,,                \\
\label{orbka}
{\hat K}_{ij}^{(a)}&=&
\frac{1}{2}\eta(\eta r_1+r_3)(1+n_{ib}n_{jb})\,.
\end{eqnarray}

The general form of spin-orbital superexchange model (\ref{HJ})
suggests that the above symmetry breaking leads indeed to an effective 
spin model with broken symmetry between magnetic interactions along 
different cubic axes. By averaging over the orbital operators one finds 
indeed different effective magnetic exchange interactions, $J_c$ along 
the $c$ axis and $J_{ab}$ within the $ab$ planes:
\begin{equation}
\label{Jij}
J_{c} =\left\langle{\hat J}_{ij}^{(c)}\right\rangle\,, \hskip 2cm
J_{ab}=\left\langle{\hat J}_{ij}^{(a)}\right\rangle\,.
\end{equation}
The interactions in the $ab$ planes could in principle still take two 
different values in case of finite lattice strain discussed below, 
making both $\{a,b\}$ axes inequivalent, but here we want just to 
point out the symmetry breaking between the $c$ axis and the $ab$ 
planes, which follows from the density distribution (\ref{nn}) 
and explains the nonequivalence of spin interactions in the $C$-AF 
phase of the $R$VO$_3$ perovskites \cite{Kha01}. 

Apart from the superexchange there are in general also interactions due 
to the couplings to the lattice that control the orbitals. In the cubic 
vanadates these interactions are expected to be weak, but nevertheless
they influence significantly the spin-orbital fluctuations and decide
about the observed properties in the $R$VO$_3$ family. We write the 
orbital interactions, $\propto\tau^z_i\tau^z_j$, induced by the 
GdFeO$_3$ distortions and by the JT distortions of the lattice using 
two parameters, $V_{ab}$ and $V_c$,
\begin{equation}
\label{HJT}
{\cal H}_V =
 V_{ab}\sum_{\langle ij\rangle\parallel c}\tau^z_i\tau^z_j
-V_c\sum_{\langle ij\rangle\parallel c}\tau^z_i\tau^z_j\,.
\end{equation}
The orbital interaction along the $c$ axis $V_c$ plays here a crucial 
role and allows one to switch between the two types of magnetic order, 
$C$-AF and $G$-AF phase \cite{Hor03}, stabilizing simultaneously either 
$G$-AO or $C$-AO order.

However, the description in terms of the GKR does
not suffice and the ground state of spin-orbital model for the $R$VO$_3$ 
perovskites, which consists of the superexchange and the effective 
orbital interactions,
\begin{equation}
\label{som} {\cal H}_{S\tau} ={\cal H}_J + {\cal H}_V\,.
\end{equation}
may also be entangled due to the quantum coupling between spin $S=1$
and orbital $\tau=1/2$ operators along the $c$ axis, see Eq.
(\ref{orbjc}). In constrast, the orbital fluctuations in the $ab$
planes are quenched due to the occupied $c$ orbitals at each site
(\ref{nn}), so spins and orbitals disentangle.
Possible entanglement between spin $({\vec S}_i\cdot {\vec S}_j)$
and orbital $({\vec\tau}_i\cdot {\vec\tau}_j)$ operators along the
bonds $\langle ij\rangle\parallel c$ in the $R$VO$_3$ perovskites, 
and the applicability of the GKR to these 
systems, may be investigated by evaluating intersite spin and 
orbital correlations (to make these two functions comparable, 
we renormalized the spin correlations by the factor $\frac14$),
\begin{eqnarray}
\label{sij}
S_{ij}&=&\frac14\langle{\vec S}_i\cdot{\vec S}_j\rangle\,, \\
\label{tij}
T_{ij}&=&\langle{\vec\tau}_i\cdot{\vec\tau}_j\rangle\,,
\end{eqnarray}
and comparing them with each other. A key quantity that measures 
spin-orbital entanglement is the composite correlation function
\cite{Ole06},
\begin{equation}
\label{cij}
C_{ij}=\frac14
\left\{\big\langle({\vec S}_i\cdot{\vec S}_j)
                ({\vec\tau}_i\cdot{\vec\tau}_j)\big\rangle
       -\big\langle{\vec S}_i\cdot{\vec S}_j   \big\rangle
      \big\langle{\vec\tau}_i\cdot{\vec\tau}_j \big\rangle\right\}\,.
\end{equation}
When $C_{ij}=0$, the spin and orbital operators are disentangled and
their MF decoupling is exact, while if $C_{ij}<0$ ---
spin and orbital operators are entangled and the MF decoupling not
justified.

\begin{figure}[t!]
\begin{center}
\begin{minipage}{0.45\textwidth}
    \includegraphics[width=\textwidth]{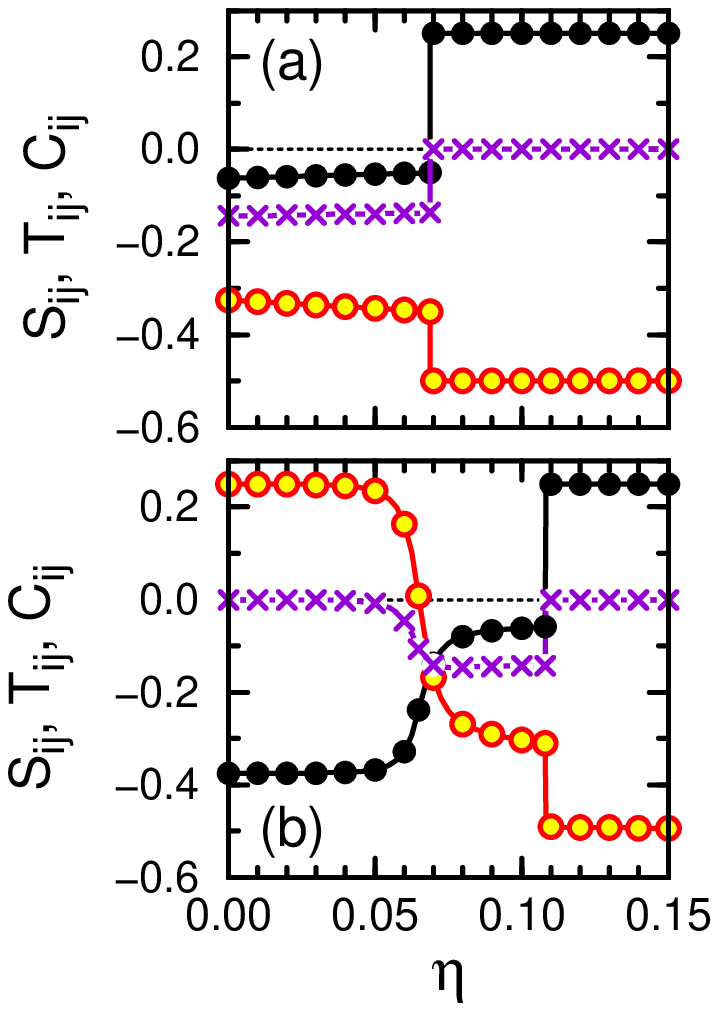}
\end{minipage}
 \quad
\begin{minipage}{0.44\textwidth}
    \includegraphics[width=\textwidth]{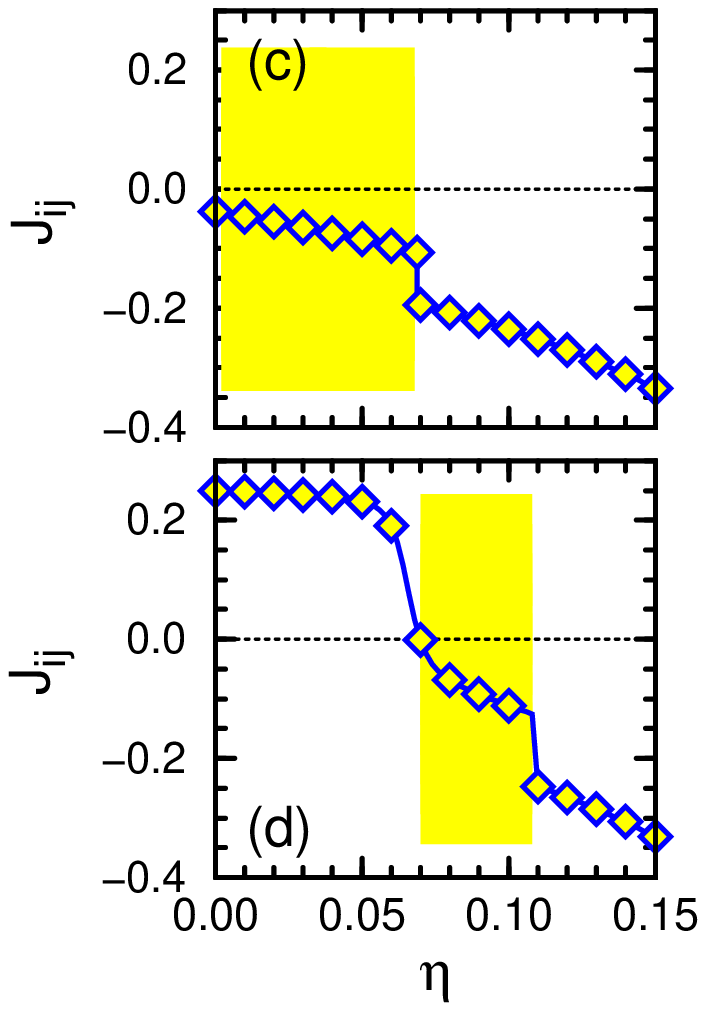}
\end{minipage}
\end{center}
\caption{
Evolution of intesite correlations and exchange constants along the
$c$ axis obtained by exact diagonalizaton of spin-orbital model on a
chain of $N=4$ sites with periodic boundary conditions, with
${\hat J}_{ij}^{(c)}$ and ${\hat K}_{ij}^{(c)}$ given by
Eqs. (\ref{orbjc}) and (\ref{orbkc}), for increasing Hund's exchange
$\eta$:
(a),(b) intersite spin $S_{ij}$ (\ref{sij}) (filled circles), 
orbital (\ref{tij}) (empty circles), and
spin-orbital $C_{ij}$ (\ref{cij}) ($\times$) correlations;
(c),(d) the corresponding spin exchange constants $J_{ij}$ (\ref{Jij}).
In the shaded areas of (c) and (d) the spin correlations $S_{ij}<0$ do
not follow the sign of the exchange constant $J_{ij}<0$, and the
classical GKR are violated.
Parameters: (a),(c) $V_c=0$, and (b),(d) $V_c=J$.
}
\label{fig:enta}
\end{figure}

The numerical results for a 1D chain along the $c$
axis described by vanadate spin-orbital model (\ref{som}) are shown in
Fig. \ref{fig:enta}. One finds entangled spin-orbital states with all
three $S_{ij}$, $T_{ij}$ and $C_{ij}$ correlations being negative in
the spin-singlet ($S=0$) regime of fluctuating $yz$ and $zx$ orbitals,
obtained for $\eta< 0.07$ [Fig. \ref{fig:enta}(a)]. Therefore, the
complementary behavior of spin (\ref{sij}) and orbital (\ref{tij})
correlations is absent in this regime of parameters and the
GKR are violated.
In addition, composite spin-orbital correlations (\ref{cij}) are here
finite ($C_{ij}<0$), so spin and orbital variables are entangled, and
the MF factorization of spin-orbital operators fails. In a similar 
$d^1$ model for the perovskite titanates (with $S=1/2$) one finds even
somewhat stronger spin-orbital entanglement and the regime of $\eta$
with $C_{ij}<0$ is broader (i.e., $\eta< 0.21$) \cite{Ole06}. At the
point $\eta=0$ one recovers then the SU(4) model with
$S_{ij}=T_{ij}=C_{ij}=-0.25$, and the ground state is an entangled
SU(4) singlet, involving a linear combination of
(spin singlet/orbital triplet) and
(spin triplet/orbital singlet) states.

To provide further evidence that the GKR do not
apply to spin-orbital model (\ref{som}) in the regime of small $\eta$,
we compare spin exchange constants $J_{ij}$ (\ref{Jij}) shown in Fig.
\ref{fig:enta}(c) with spin correlations $S_{ij}$ (\ref{sij}), see
Fig. \ref{fig:enta}(a). One finds that exchange interaction is formally
FM ($J_{ij}<0$) in the orbital-disordered phase in the regime of
$\eta<0.07$, but it is accompanied by AF spin correlations ($S_{ij}<0$).
Therefore $J_{ij}S_{ij}>0$ and the ground state energy would be enhanced
in an ordered state, when calculated in the MF decoupling of 
spin-orbital operators \cite{Ole06}. This at first instance somewhat 
surprising result is a consequence of {\it `dynamical'\/} nature of 
exchange constants $\hat{J}_{ij}^{(c)}$ which exhibit large fluctuations 
\cite{Ole06}, measured by the second moment,
$\delta J=\{\langle (\hat{J}_{ij}^{(\gamma)})^2\rangle-J_{ij}^2\}^{1/2}$.
For instance, in $d^2$ model (\ref{som}) the orbital bond correlations
change dynamically from singlet to triplet, resulting in large
$\delta J=\frac{1}{4}\{1-(2T_{ij}+\frac{1}{2})^2\}^{1/2}\simeq 0.247$,
i.e., $\delta J>|J_{ij}|$.

Remarkably, finite spin-orbital correlations $C_{ij}<0$ and similar
violation of the GKR are found also at finite
orbital interaction (\ref{HJT}) induced by the lattice, $V_c>0$.
Representative results obtained for $V_c=J$ are shown in Figs.
\ref{fig:enta}(b) and \ref{fig:enta}(d). At small $\eta$ FO order is 
induced, and in this regime the GKR are followed 
by the AF/FO phase (similar to the FM/AO phase at large $\eta$ which
also follows the GKR). However, for intermediate
Hund's exchange $\eta\sim 0.07$ FO order is destabilized and the
entangled AF/AO phase appears, with similar spin, orbital and
composite spin-orbital coorrelations as found before at $V_c=0$
and $\eta=0$ [Figs. \ref{fig:enta}(a)]. Also in this case FM exchange
($J_{ij}<0$) coexists with AF spin correlations ($S_{ij}<0$).
Thus we conclude that orbital interactions induced by the lattice modify 
the regime of entangled spin-orbital states in the intermediate AF/AO 
phase which may be moved to more realistic values of $\eta\sim 0.1$, 
and cannot eliminate it completely. In addition, the transition between 
the FO/AF and AO/AF phase is {\it continuous\/} \cite{Ole06} due to 
the structure of orbital superexchange which contains terms
(\ref{exo1}) responsible for non-conservation of orbital quantum
numbers.

\section{Experimental evidence of orbital fluctuations in LaVO$_3$/YVO$_3$}
\label{sec:rvo}

Before discussing the exotic magnetic properties and the phase diagram 
of the $R$VO$_3$ perovskites, we will consider the influence of 
magnetism on the optical spectra of LaVO$_3$, starting with a general 
formulation of the theory. While exchange constants may be extracted 
from the spin-orbital superexchange model (\ref{Jij}), it is frequently 
not realized that virtual charge excitations that contribute to the 
superexchange are responsible as well for the optical absorption, 
thus the superexchange and the optical 
absorption are intimately related to each other via the optical sum 
rule \cite{Bae86}. This is not so surprising as when electrons are 
almost localized in a Mott insulator, the only kinetic energy which is 
left and decides about the optical spectral weight is associated with 
virtual excitations contributing to superexchange. Therefore, in Mott 
insulators the thermal evolution of optical spectral weight can be 
deduced from the superexchange \cite{Aic02}. In a system with orbital 
degeneracy the optical spectra consist of several multiplet transitions, 
and the kinetic energy $K_n^{(\gamma)}$ (due to $d-d$ excitations) 
associated with each of them can be determined from the superexchange 
(\ref{HJ}) using the Hellman-Feynman theorem \cite{Kha04},
\begin{equation}
\label{hefa} K_n^{(\gamma)}=2\left\langle
H_n^{(\gamma)}(ij)\right\rangle\,.
\end{equation}
Note that $K_n^{(\gamma)}$ is negative and corresponds to the $n$'th 
multiplet state of the transition metal ion, created by a charge 
excitation along a bond $\langle ij\rangle\parallel\gamma$. It is 
obvious that the thermal excitation values $\langle\cdots\rangle$
depend sensitively on the magnetic structure, i.e., whether spin 
correlations on a bond $\langle ij\rangle$ are FM or AF.

Thus it is natural to decompose the optical sum rule which is usually 
formulated in terms of the {\it total\/} kinetic energy for polarization 
$\gamma$,
\begin{equation}
\label{opsatot}
K^{(\gamma)}=2J\sum_n\big\langle H_n^{(\gamma)}(ij)\big\rangle,
\end{equation}
into {\it partial optical sum rules\/} 
for individual Hubbard subbands \cite{Kha04},
\begin{equation}
\label{opsa}
\frac{a_0\hbar^2}{e^2}\int_0^{\infty}\sigma_n^{(\gamma)}(\omega)
d\omega=-\frac{\pi}{2}K_n^{(\gamma)}=-\pi\left\langle
H_n^{(\gamma)}(ij)\right\rangle\,,
\end{equation}
where $\sigma_n^{(\gamma)}(\omega)$ is the contribution of band
$n$ to the optical conductivity for polarization along the
$\gamma$ axis, $a_0$ is the distance between transition metal
ions, and the tight-binding model with nearest neighbor hopping is
assumed. Equation (\ref{opsa}) provides a practical way of
calculating the optical spectral weights from spin-orbital
superexchange models, such as the one derived for the $R$VO$_3$
perovskites (\ref{som}). Note that the total optical intensity 
(\ref{opsatot}) is of less interest here as it has a much weaker 
temperature dependence and does not allow one for a direct insight 
into the nature of the electronic structure. In addition, it might 
be also more difficult to resolve from experiment.

In order to apply the above theory to the $R$VO$_3$ perovskites, we 
write the superexchange operator $H^{(\gamma)}(ij)$ for a bond 
$\langle ij\rangle\parallel\gamma$, contributing to operator 
${\cal H}_J$ (\ref{HJ}), as a superposition of 
$d_i^2d_j^2\rightleftharpoons d_i^3d_j^1$ charge excitations to 
different upper Hubbard subbands labelled by $n$ \cite{Kha04},
\begin{equation}
\label{Hn} H^{(\gamma)}(ij)=\sum_n H_n^{(\gamma)}(ij)\,.
\end{equation}
One finds the superexchange terms $H^{(c)}_n(ij)$ for a bond
${\langle ij\rangle}$ along the $c$ axis \cite{Kha04},
\begin{eqnarray}
\label{H1c} H_1^{(c)}(ij)&=&-\frac{1}{3}Jr_1 (2+\vec
S_i\!\cdot\!\vec S_j)
\left(\frac{1}{4}-\vec \tau_i\cdot\vec \tau_j\right)\,,                  \\
\label{H2c} H_2^{(c)}(ij)&=&-\frac{1}{12}J(1-\vec S_i\!\cdot\!\vec
S_j) \left(\frac{7}{4}-\tau_i^z\tau_j^z-\tau_i^x\tau_j^x
+5\tau_i^y \tau_j^y\right)\,,                                           \\
\label{H3c} H_3^{(c)}(ij)&=&-\frac{1}{4}Jr_3 (1-\vec
S_i\!\cdot\!\vec S_j)
\left(\frac{1}{4}+\tau_i^z\tau_j^z+\tau_i^x\tau_j^x -\tau_i^y
\tau_j^y\right)\,,
\end{eqnarray}
and $H^{(ab)}_n(ij)$ for a bond in the $ab$ plane,
\begin{eqnarray}
\label{H1a} H_1^{(ab)}(ij)&=&-\frac{1}{6}Jr_1\left(2+\vec
S_i\!\cdot\!\vec S_j\right)
\left(\frac{1}{4}-\tau_i^z\tau_j^z\right)\,,                 \\
\label{H2a} H_2^{(ab)}(ij)&=&-\frac{1}{8}J\left(1-\vec
S_i\!\cdot\!\vec S_j\right) \left(\frac{19}{12}\mp
\frac{1}{2}\tau_i^z
\mp \frac{1}{2}\tau_j^z-\frac{1}{3}\tau_i^z\tau_j^z\right)\,,           \\
\label{H3a} H_3^{(ab)}(ij)&=&-\frac{1}{8}Jr_3\left(1-\vec
S_i\!\cdot\!\vec S_j\right) \left(\frac{5}{4}\mp \frac{1}{2}\tau_i^z
\mp \frac{1}{2}\tau_j^z+\tau_i^z\tau_j^z\right)\,.
\end{eqnarray}
These expressions show that the spin correlations along the $c$
axis and within the $ab$ planes,
\begin{equation}
\label{spins}
   s_c=\langle\vec{S}_i\cdot\vec{S}_j\rangle_{c}\,,   \hskip 1cm
s_{ab}=\langle\vec{S}_i\cdot\vec{S}_j\rangle_{ab}\,,
\end{equation}
as well as the orbital correlations, play an important role in the
intensity distribution in optical spectroscopy. From the form of 
the above superexchange contributions one sees that high-spin 
excitations $H_1^{(\gamma)}(ij)$ support the FM coupling while 
the low-spin ones, $H_2^{(\gamma)}(ij)$ and $H_3^{(\gamma)}(ij)$,
contribute with AF couplings.

We have determined the exchange constants in LaVO$_3$ by averaging
over the orbital operators, see Eqs. (\ref{Jij}). The case of the
$ab$ planes is straightforward as only the average densities 
$\langle n_{ia}\rangle$ and $\langle n_{ib}\rangle$ are needed to 
determine $J_{ab}$, and at large $\eta$ they follow from the $G$-AO 
order in these planes. At $\eta=0$ the orbital correlations along the 
$c$ axis result from orbital fluctuations in the 1D orbital chain. In 
this limit the orbital correlations are the same as for the AF 
Heisenberg chain, i.e., 
$\langle{\vec\tau}_i\cdot{\vec\tau}_j\rangle=-0.4431$ and the ground 
state is disordered, with $\langle\tau_i^z\rangle=0$. Nevertheless, for 
this disordered state the result for $J_{ab}$ is similar as for the 
$G$-AO phase \cite{Ole07}. 

\begin{figure}[t!]
\sidecaption[t]
\includegraphics[width=7.0cm]{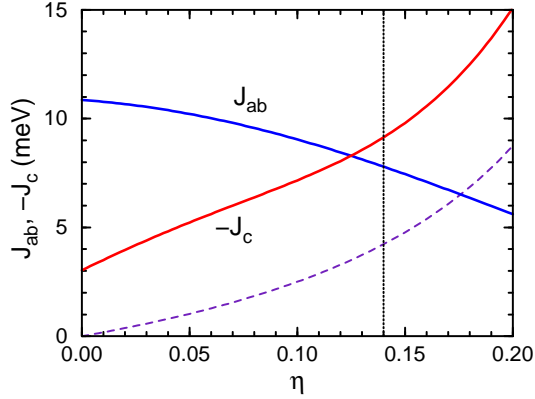}
\caption{ Exchange constants $J_{ab}$ and $-J_c$ (\ref{Jij}) 
calculated from Eqs. (\ref{orbja}) and (\ref{orbjc}) in the
$C$-AF phase of LaVO$_3$ for increasing $\eta$ (solid lines).
Dashed line shows the value of $-J_c$ obtained for classical
orbital order (\ref{Jccaf}) according to GKR,
$\langle\vec\tau_i\cdot\vec\tau_j\rangle=-\frac14$. 
A representative value of $\eta=0.14$ (for $U=5.0$ and
$J_H=0.7$ eV) is marked by dotted line. Parameters of the
model (\ref{som}): $J=35$ meV, $V_c=V_{ab}=0$. }
\label{fig:lavo}
\end{figure}

For the disordered (fluctuating) $\{a,b\}$ orbital state at $\eta=0$, 
the AF exchange interactions in $ab$ planes (see Fig. \ref{fig:lavo})
result solely from singly occupied $c$ orbitals (\ref{nn}), which
are active in these planes and contribute by their double
occupancies in excited state with AF superexchange.  One expects
that the exchange constants along the $c$ axis in the $C$-AF phase
could be deduced from Eqs. (\ref{Jij}), as spin and orbital order
are complementary \cite{Miy06}. It is quite remarkable that at the
same time finite FM interactions $-J_c\simeq 3$ meV are obtained
at $\eta=0$ (Fig. \ref{fig:lavo}). They follow from the orbital
fluctuations which dominate at low values of $\eta$.
This mechanism of FM exchange adds to the one known in systems with
real orbital order at finite $\eta$ --- the latter mechanism gradually 
takes over when $\eta$ increases and the $G$-AO order develops and 
reduces the orbital fluctuations. At finite $\eta>0$ we used the linear
orbital-wave theory \cite{vdB99} to determine the intersite orbital 
correlations $\langle{\vec\tau}_i\cdot{\vec\tau}_j\rangle$ and the order 
parameter $\langle\tau_i^z\rangle$, for more details see Ref. 
\cite{Ole07}. At $\eta=0.14$ representative for LaVO$_3$, the FM 
interactions are stronger than from AF ones, $|J_c|>J_{ab}$. Indeed, 
this early prediction of the theory \cite{Kha01} agrees qualitatively 
with larger average FM exchange $J_c<0$ in the $C$-AF phase of YVO$_3$ 
than the AF exchange $J_{ab}>0$ in the $ab$ planes, see below.

We emphasize that the strong FM exchange along the $c$ axis
follows from the orbital fluctuations, and the rigid $G$-AO order
obtained in the limit of strong orbital interactions $\{V_{ab},V_c\}$
(\ref{HJT}) would give a much weaker FM interaction,
\begin{equation}
\label{Jccaf} J_c^{G-{\rm AO}}=-\frac{1}{2}\eta r_1J\,,
\end{equation}
see Fig. \ref{fig:lavo}. The FM interaction $J_c^{G-{\rm AO}}$
(\ref{Jccaf}) vanishes at $\eta=0$, is triggered by finite Hund's
exchange $\eta$ and increases in lowest order linearly with
$\eta$. This behavior follows the conventional mechanism of FM
interactions induced by finite Hund's exchange in the states
with AO order, as for instance in KCuF$_3$ \cite{Fei97} or in
LaMnO$_3$ \cite{Fei99}.

\begin{figure}[t!]
\sidecaption[t]
\includegraphics[width=7.5cm]{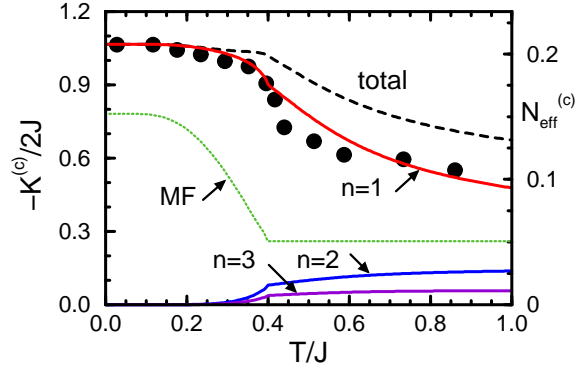}
\caption{ Kinetic energy $K_n^{(c)}$ (solid lines) for the optical
subband $n$ and total $K^{(c)}$ (dashed line)
obtained from the spin-orbital model (\ref{som}). Filled circles
show the effective carrier number $N_{eff}^{(c)}$ (in the energy 
range $\omega<3$ eV) for LaVO$_3$, presented in Fig. 5 of Ref. 
\cite{Miy02}. Dotted line shows
$K_1^{(c)}$ obtained from the MF decoupling (\ref{wc21}).
Parameters: $\eta=0.12$, $V_c=0.9J$, $V_{ab}=0.2J$. }
\label{fig:osw}
\end{figure}

A crucial test of the present theory which demonstrates that orbital
fluctuations are indeed present in LaVO$_3$, concerns the
temperature dependence of the low-energy (high-spin) spectral
weight in optical absorption along the $c$ axis $-K_1^{(c)}/2J$. 
According to experiment \cite{Miy02} it decreases by about 50\% 
between low temperature and $T=300$ K. In contrast, the result obtained 
by averaging the high-spin superexchange term $H_1^{(c)}(ij)$ 
(\ref{H1c}) for polarization along the $c$ axis assuming robust 
$G$-AO order is,
\begin{equation}
\label{wc21}
w_{c1}^{G-{\rm AO}}=\frac{1}{6}r_1\big(s_c+2\big)\,,
\end{equation}
where the spin correlation function $s_c$ (\ref{spins}) is 
responsible for the entire temperature dependence of the low-energy 
spectral weight. Equation (\ref{wc21}) 
predicts decrease of $w_{c1}$ of only about 27\%, see Fig.
\ref{fig:osw}, and the maximal possible reduction of $K_1^{(c)}$ 
reached at $s_c=0$ in the limit of $T\to\infty$ is by 33\%. This 
result proves that the scenario with frozen $G$-AO order in LaVO$_3$ 
is {\it excluded by experiment\/} \cite{Miy05}.

In contrast, when a cluster method which allows to include orbital 
fluctuations along the $c$ axis is used to determine the optical 
spectral weight from the high-spin superexchange term 
(\ref{H1c}) \cite{Kha04}, the temperature dependence resulting from 
the theory follows the experimental data \cite{Miy02}. This may be
considered as a remarkable success of the theory based on the
spin-orbital superexchange model derived for the $R$VO$_3$
perovskites.

However, the experimental situation in the cubic vanadates is more
complex and full of puzzles. One is connected with the second magnetic
transition in YVO$_3$, as we already mentioned in Sec.
\ref{sec:som}. The magnetic transition at $T_{N2}=77$ K is
particularly surprising as the staggered moments are approximately 
parallel to the $c$ axis in the $G$-AF phase, and rotate above $T_{N2}$ 
to the $ab$ planes in the $C$-AF phase, with some small alternating 
$G$-AF component along the $c$ axis \cite{Ren00}.
While the orientation of spins in $C$-AF and $G$-AF phase follow in a
straightforward manner from the model, i.e., are consistent with the
expected anisotropy due to spin-orbit coupling \cite{Hor03}, the 
observed magnetization reversal with the weak FM component remains
puzzling. Therefore, in spite of the suggested mechanism based on the 
entropy increase in the $C$-AF phase \cite{Ole07}, the lower magnetic 
transition in YVO$_3$ remains mysterious. Secondly, the scale of 
magnetic excitations is considerably reduced for the $C$--AF phase 
(by a factor close to two) as compared with the exchange constants
deduced from magnons measured in the $G$-AF phase \cite{Ulr03}. In
addition, the magnetic order parameter in the $C$-AF phase of
LaVO$_3$ is strongly reduced to $\simeq 1.3\mu_B$, which cannot be
explained by the quantum fluctuations in the $C$-AF phase (being
only 6\% for $S=1$ spins \cite{Rac02}). Finally, the $C$-AF phase
of YVO$_3$ is dimerized. Until now, only this last feature found a
satisfactory explanation in the theory \cite{Sir03,Sir08}, see
below.

\begin{figure}[t!]
\sidecaption[t]
\includegraphics[width=7.5cm]{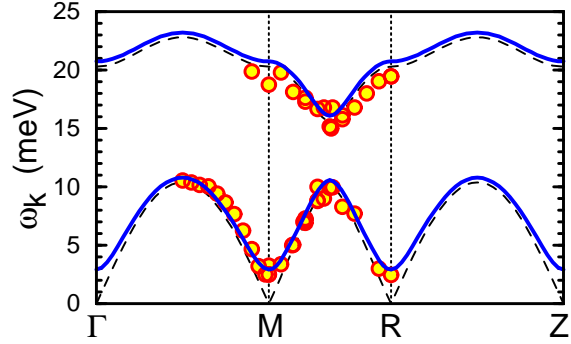}
\caption{Spin-wave dispersions $\omega_{\bf k}$ obtained in the
LSW theory (\ref{spinw}) for the $C$-AF phase of YVO$_3$ (lines),
and measured by neutron scattering at $T=85$ K \cite{Ulr03} (circles).
Parameters: $J_{ab}=2.6$ meV, $J_c=-3.1$ meV, $\delta_s=0.35$, and
$K_z=0.4$ eV (full lines), $K_z=0$ (dashed lines). The high symmetry 
points are: $\Gamma=(0,0,0)$, $M=(\pi,\pi,0)$, $R=(\pi,\pi,\pi)$, 
$Z=(0,0,\pi)$. }
\label{fig:swcafd}
\end{figure}

We remark that the observed dimerization in the magnon dispersions
may be seen as a signature of {\it entanglement in excited states\/} 
which becomes active at finite temperature. The microscopic reason of 
the anisotropy in the exchange constants
${\cal J}_{c1}\equiv{\cal J}_c(1+\delta_s)$ and 
${\cal J}_{c2}\equiv{\cal J}_c(1-\delta_s)$ 
is the tendency of the orbital chain to dimerize, activated by thermal 
fluctuations in the FM spin chain \cite{Sir08} which support dimerized 
structure in the orbital sector. As a result one finds alternating 
stronger $\propto {\cal J}_c(1+\delta_s)$ and weaker $\propto {\cal
J}_c(1-\delta_s)$ FM bonds along the $c$ axis in the dimerized
$C$-AF phase (with $\delta_s>0$). The observed spin waves may be 
explained by the following effective spin Hamiltonian for this phase 
(assuming again that the spin and orbital operators may be disentangled 
which is strictly valid only at $T=0$):
\begin{equation}
\label{hcafd}
{\cal H}_{s}={\cal J}_{c}\sum_{\langle i,i+1\rangle\parallel c}
   \left[1+(-1)^i\delta_s\right]{\vec S}_{i}\cdot{\vec S}_{i+1}
   +{\cal J}_{ab}^C\sum_{\langle ij\rangle\parallel ab}
      {\vec S}_i\cdot{\vec S}_j
   +K_z\sum_i\left(S_i^z\right)^2\,.
\end{equation}
Following the linear spin-wave theory \cite{Ole07}, 
the magnon dispersion is given by
\begin{equation}
\label{spinw} \omega_{\pm}({\bf k})= 2\left\{\left[2{\cal
J}_{ab}+|{\cal J}_c|+\frac12 K_z \pm {\cal J}_c\eta_{\bf
k}^{1/2}\right]^2 -\big(2{\cal J}_{ab}\gamma_{\bf
k}\big)^2\right\}^{1/2}\,,
\end{equation}
with
\begin{eqnarray}
\label{gamma}
\gamma_{\bf k}&=&\frac12\left(\cos_kx+\cos k_y\right)\;,  \\
\label{eta}
\eta_{\bf k}&=&\cos^2k_z+\delta_s^2\sin^2k_z\,.
\end{eqnarray}
For the numerical evaluation of figure \ref{fig:swcafd} we have used
the experimental exchange interactions \cite{Ulr03}: ${\cal
J}_{ab}=2.6$ meV, ${\cal J}_c=-3.1$ meV, $\delta_s=0.35$. Indeed,
large gap is found between two modes halfway in between the $M$
and $R$ points, and between the $Z$ and $\Gamma$ points (not
shown). Two modes measured by neutron scattering \cite{Ulr03} 
(see also figure 1) and
obtained from the present theory in the unfolded Brillouin zone
are well reproduced by the dimerized FM exchange couplings in spin
Hamiltonian (\ref{spinw}). We note that a somewhat different 
Hamiltonian with more involved interactions was introduced in 
ref. \cite{Ulr03}, but the essential features seen in the experiment 
are reproduced already by the present model $H_s$ with a single ion 
anisotropy term $\propto K_z$.

\begin{figure}[t!]
\sidecaption[t]
\includegraphics[width=7.5cm]{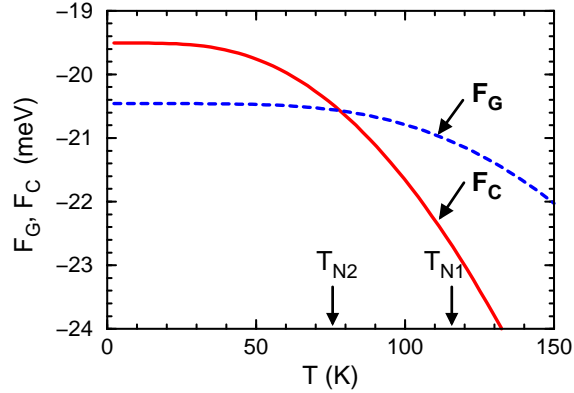}
\caption{
Free energies ${\cal F}_C$ ($C$-AF, solid line) and ${\cal F}_G$
($G$-AF, dashed line) as obtained for the spin-orbital model
(\ref{som}) using the experimental values of magnetic exchange
constants in both phases \cite{Ulr03}. The experimental magnetic
transition temperatures,
$T_{N2}\simeq 77$ K and
$T_{N1}\simeq 116$ K, are indicated by arrows.
Parameters: $J=40$ meV, $\eta=0.13$, $V_a=0.30J$, $V_c=0.84J$.
}
\label{fig:free}
\end{figure}

As the transition between the two magnetic phases, $G$-AF and
$C$-AF phase, occurs in YVO$_3$ at finite temperature, the entropy
has to play an important role. As mentioned above, the exchange 
constants found in the $C$-AF phase of YVO$_3$ (Fig. \ref{fig:swcafd}) 
are considerably lower than the corresponding values in the $G$-AF 
phase, $J_{ab}=J_c\simeq 5.7$ meV \cite{Ulr03}. 
As a result of weaker exchange interactions, the spin  
entropy of the $C$ phase will grow faster than that of the $G$ phase, 
and induce the $G\rightarrow C$ transition. However, starting from our 
model (\ref{som}) we do not find this strong reduction of energy scale 
in the $C$-AF phase. Other mechanism like the fluctuation of $n_{xy}$ 
occupancy has been invoked to account for this reduction \cite{Ole07}.
Here we will simply adopt the experimental values for the exchange
constants in the $C$-AF phase.

Using linear spin-wave and orbital-wave theory, the spin and orbital 
entropy normalized per one vanadium ion was calculated and compared 
for both magnetic phases of YVO$_3$ \cite{Ole07}. Using the 
experimental parameters \cite{Ulr03} one finds that: 
 ($i$) the entropy ${\cal S}_C$ for the $C$-AF phase is larger that 
     ${\cal S}_G$ for the $G$-AF phase, and 
($ii$) the spin entropy grows significantly faster with temperature 
     than the orbital entropy for each phase.      
Therefore, we conclude that the spin entropy gives here a leading
contribution and is responsible for a fast decrease of the free 
energy in the $C$-AF phase which is responsible for the observed
magnetic transition at $T_{N2}$ \cite{Ole07}, see Fig.~\ref{fig:free}.

\section{Orbital and magnetic transition in the $R$VO$_3$ perovskites}
\label{sec:phd}

\subsection{Spin-orbital-lattice coupling}
\label{subsec:sol}

Experimental studies have shown that the $C$-AF order is common to the 
entire family of the $R$VO$_3$ vanadates, where $R$=Lu,$\cdots$,La. In
general the structural (orbital) transition occurs first. i.e., 
$T_{N1}<T_{\rm OO}$, except for LaVO$_3$ with $T_{N1}\simeq T_{\rm OO}$
\cite{Miy03,Miy06}. When the ionic radius $r_R$ decreases, the N\'eel 
temperature $T_{N1}$ also decreases, while the orbital transition 
temperature $T_{\rm OO}$ increases first, passes through a maximum 
close to YVO$_3$, and decreases afterwards when LuVO$_3$ is approached.
Knowing that quantum fluctuations and spin-orbital entanglement play
so important role in the perovskite vanadates, it is of interest to
ask whether the spin-orbital model (\ref{som}) 
is able to describe this variation of
$T_{\rm OO}$ and $T_{N1}$ with decreasing radius $r_R$ of $R$ ions in
$R$VO$_3$ \cite{Miy03}. It is clear that the nonmonotonic dependence of
$T_{\rm OO}$ on $r_R$ cannot be reproduced just by the superexchange,
as a maximum in $T_{\rm OO}$ requires two mechanisms which oppose each 
other. In fact, the decreasing V--O--V angle ($\Theta$ along the $c$ 
axis) with decreasing ionic radius $r_R$ along the $R$VO$_3$ perovskites
\cite{Ren03,Ree06,Sag06,Sag07} reduces somewhat both the hopping $t$ and 
superexchange $J$ (\ref{sex}), but we shall ignore this effect here and 
concentrate ourselves on the leading dependence 
on orbital correlations which are controlled by lattice distortions.

The model introduced in Ref. \cite{Hor08} to describe the phase diagram 
of $R$VO$_3$ includes the spin-orbital-lattice coupling by the terms:
($i$) the superexchange $H_J$ (\ref{HJ}) between $V^{3+}$ ions in the 
$d^2$ configuration with $S=1$ spins \cite{Kha01},
($ii$) intersite orbital interactions $H_V$ (\ref{HJT}) (which
originate from the coupling to the lattice and play an important role
in the transition between the $C$-AF and $G$-AF phase), 
($iii$) the crystal-field splitting $\propto E_z$ between $yz$ and
$zx$ orbitals, and
($iv$) orbital-lattice term $\propto gu$ which induces orbital 
polarization when the lattice strain (distortion) $u$ inreases.
The Hamiltonian consists thus of several terms \cite{Hor08},
\begin{equation}
\label{rvo3}
{\cal H}={\cal H}_{J}+{\cal H}_V(\vartheta)+E_z(\vartheta)
\sum_i\!e^{i{\vec R}_i{\vec Q}}\tau_i^z
-gu\sum_i\tau_i^x+\frac12 N K\{u-u_0(\vartheta)\}^2\,.
\end{equation}
Except for the superexchange ${\cal H}_J$ (\ref{HJ}), all the other 
terms in Eq. (\ref{rvo3}) depend on the tilting angle $\vartheta$, which 
we use to parameterize the $R$VO$_3$ perovskites below. It is related to 
the V--O--V angle $\Theta=\pi-2\vartheta$, which decreases with 
increasing ionic radius $r_R$ ($\Theta=180^{\circ}$ corresponds to an 
ideal perovskite structure). By analyzing the structural data of the 
$R$VO$_3$ perovskites \cite{Ren03,Ree06,Sag06,Sag07} one arrives at the 
following empirical relation between $r_R$ and $\vartheta$:
\begin{equation}
\label{rR}
r_R=r_0-\alpha\,\sin^22\vartheta\,,
\end{equation}
with $r_0=1.5$ \AA{} and $\alpha=0.95$ \AA{}.

The crystal-field splitting of $\{yz,zx\}$ orbitals ($E_z>0$) alternates 
in the $ab$ planes and is uniform along the $c$ axis, with a modulation 
vector ${\vec Q}=(\pi,\pi,0)$ in cubic notation --- it supports the 
$C$-AO order, and not the observed (weak) $C$-AO order. The orbital 
interactions induced by the distortions of the VO$_6$ octahedra and by 
GdFeO$_3$ distortions of the lattice, $V_{ab}>0$ and $V_c>0$, also 
favor the $C$-AO order (like $E_z>0$). 
The orbital interaction $V_c$ counteracts the orbital superexchange 
$\propto J$ (\ref{orbkc}), and has only rather weak dependence on
$\vartheta$, so it suffices to choose a constant $V_c=0.26J$ to
reproduce an almost simultaneous onset of spin and orbital order in 
LaVO$_3$, with $T_{\rm OO}\simeq T_{N1}$, as observed \cite{Miy03}.
One finds $T_{N1}^{\rm exp}=147$ K taking $J=200$ K in the present 
model (\ref{rvo3}), which reproduces well the experimental value
$T_{N1}^{\rm exp}=143$ K for LaVO$_3$ \cite{Miy03}.

The last two terms in Eq. (\ref{rvo3}) descibe the orbital-lattice 
coupling via the orthorhombic strain $u=(b-a)/a$, where $a$ and $b$ are 
the lattice parameters of the $Pbnm$ structure, $K$ is the force 
constant, and $N$ is the number of $V^{3+}$ ions. Unlike $E_z$, the 
coupling $gu>0$ acts as a transverse field in the pseudospin space and 
favors that one of the
two linear combinations $\frac{1}{\sqrt{2}}(|a\rangle_i\pm|b\rangle_i)$ 
of active $t_{2g}$ orbitals is occupied at site $i$. By minimizing
the energy over $u$, one finds 
\begin{equation}
\label{geffT}
g_{\rm eff}(\vartheta;T)\equiv gu(\vartheta;T)
=gu_0(\vartheta)+\frac{g^2}{K}\langle\tau^x\rangle_T\,,
\end{equation}
which shows that the global distortion $u(\vartheta;T)$ consists of
($i$) a pure lattice contribution $u_0(\vartheta)$, and
($ii$) a contribution due the orbital polarization
$\propto\langle\tau^x\rangle$ which is determined self-consistently.

\subsection{Dependence on lattice distortion}
\label{subsec:la}

In order to investigate the phase diagram of the $R$VO$_3$ perovskites
one needs still information on the functional dependence of the 
parameters $\{E_z,V_{ab},g_{\rm eff}\}$ of the microscopic model 
(\ref{rvo3}) on the tilting angle $\vartheta$. The GdFeO$_3$-like 
distortion is parametrized by two angles $\{\vartheta,\varphi\}$ 
describing rotations around the $b$ and $c$ cubic axes, as explained in 
Ref. \cite{Pav05}. 
Here we adopted a representative value of $\varphi=\vartheta/2$,
similar as in the perovskite titanates. Therefore, we used only a single
rotation angle $\vartheta$ in Eq. (\ref{rvo3}), which is related to the
ionic size by Eq. (\ref{rR}). Functional dependence of the crystal-field
splitting $E_z\propto\sin^3\vartheta\cos\vartheta$ on the angle 
$\vartheta$ may be derived from the point charge model \cite{Hor08}, 
using the structural data for the $R$VO$_3$ perovskites 
\cite{Ren03,Ree06,Sag06,Sag07}. It is expected that the functional 
dependence of $V_{ab}$ follows the crystal-field term, so we write:
\begin{eqnarray}
\label{Ez}
E_z(\vartheta)&=&J\,v_z\,\sin^3\vartheta\cos\vartheta\,, \\
\label{vab}
V_{ab}(\vartheta)&=&J\,v_{ab}\,\sin^3\vartheta\cos\vartheta.
\end{eqnarray}
Qualitatively, increasing $E_z$ and $V_{ab}$ with increasing lattice
distortion and tilting angle $\vartheta$ do favor the orbital order, 
so the temperature $T_{\rm OO}$ is expected to increase.

A maximum observed in the dependence of $T_{\rm OO}$ on $r_R$ (or 
$\vartheta$) may be reproduced within the present model (\ref{rvo3})
only when a competing orbital polarization interaction 
$g_{\rm eff}(\vartheta;T)$ (\ref{geffT}) increases faster with 
$\vartheta$ when the ionic radius $r_R$ is reduced than
$\{E_z,V_{ab}\}$. Both $u_0$ and $\langle\tau^x\rangle$ in Eq. 
(\ref{geffT}) are expected to increase with increasing tilting angle 
$\vartheta$. Below we present the results obtained with a semiempirical 
relation,
\begin{equation}
\label{geff}
g_{\rm eff}(\vartheta)=J\,v_{g}\,\sin^5\vartheta\cos\vartheta\,,
\end{equation}
as postulated in Ref. \cite{Hor08}. Altogether, model (\ref{som}) 
depends on three parameters: $\{v_z,v_{ab},v_g\}$ which could be 
selected \cite{Hor08} to reproduce the observed dependence of
orbital and magnetic transition temperature on the ionic radius 
$r_R$ in the $R$VO$_3$ perovskites, see below.

\subsection{Evolution of spin and orbital order in $R$VO$_3$ }
\label{subsec:evo}

Hamiltonian (\ref{rvo3}) poses a many-body problem which includes
an interplay between spin, orbital, and lattice degrees of freedom. 
A standard approach to investigate the onset of spin and orbital
order is to use the mean-field (MF) theory with on-site order 
parameters $\langle S^z\rangle$ (corresponding to $C$-AF phase) and
\begin{equation}
\label{taug}
\langle\tau^z\rangle_G\equiv\frac12
\left|\langle\tau^z_i-\tau^z_j\rangle\right|\,, 
\end{equation}
as well as the coupling between them which modifies the MF equations, 
similar to the situation encountered in the Ashkin-Teller model
\cite{Dit80}. This approach was successfully implemented
to determine the orbital and magnetic transition temperature, 
$T_{\rm OO}$ and $T_{N1}$ in LaMnO$_3$ \cite{Fei99}. It was also
applied to the $R$VO$_3$ perovskites \cite{Silva} to demonstrate that
either spin or orbital order may occur first at decreasing temperature,
depending on the amplitude of hopping parameters. However, Such a MF 
approach uses only on-site order parameters and cannot suffice when 
orbital fluctuations also contribute, e.g. stabilizing the $C$-AF phase
in LaVO$_3$ \cite{Kha01} --- then it becomes essential to determine 
self-consistently the above on-site order parameters together with 
orbital singlet correlations (\ref{tij}) on the bonds 
$\langle ij\rangle\parallel c$. The simplest approach which allows us to
determine these correlations is a cluster MF theory for a bond coupled 
to effective spin and orbital symmetry breaking fields which originate 
from its neighbors in an ordered phase. The respective transition 
temperatures are obtained when $\langle S^z\rangle>0$ 
($\langle S^z\rangle=0$) for $T<T_{N1}$ ($T>T_{N1}$), and 
$\langle\tau^z\rangle_G>0$ ($\langle\tau^z\rangle_G=0$) for 
$T<T_{\rm OO}$ ($T>T_{\rm OO}$).

\begin{figure}[t!]
\sidecaption[t]
\includegraphics[width=7.0cm]{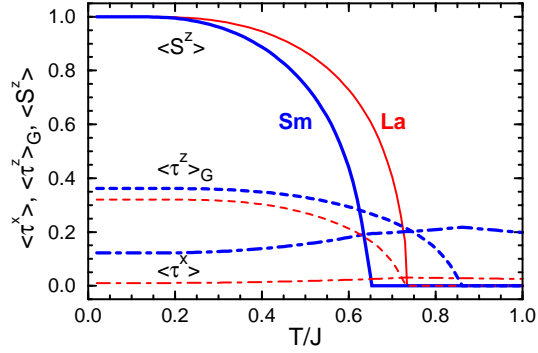}
\caption{
The orbital polarization $\langle\tau^x\rangle$ (dashed-dotted lines),
$G$-type orbital order parameter $\langle\tau^z\rangle_G$ (\ref{taug})
(dashed lines), and spin order parameter $\langle S^z\rangle$ 
(solid lines) for LaVO$_3$ and SmVO$_3$ (thin and heavy lines).
Parameters: $V_{c}=0.26J$, $v_z=17$, $v_{ab}=22$, $v_{g}=740$.
}
\label{fig:ops}
\end{figure}

Making a proper selection of the model parameters $\{v_z,v_{ab},v_g\}$
one is able to reproduce the experimental phase diagram for the onset
of $G$-AO and $C$-AF order in the $R$VO$_3$ family in the entire
range of $r_R$, see below. We start with presenting an example of the
orbital and spin phase transition in LaVO$_3$ and in SmVO$_3$, see Fig.
\ref{fig:ops}. By selecting $V_{c}=0.26J$ both $G$-AO and $C$-AF order
occur simultaneuosly in LaVO$_3$ below $T_{\rm OO}=T_{N1}\simeq 0.73J$.
The crystal field splitting $E_z$, orbital interaction $V_{ab}$, and the 
coupling to the lattice $g_{\rm eff}$ are rather small and do not 
influence the order in LaVO$_3$. We emphasize that orbital correlations 
along the $c$ axis are here practically as in the AF Heisenberg chain,
$\langle{\vec\tau}_i\cdot{\vec\tau}_j\rangle\simeq -0.44$, and the 
orbital order is considerably reduced, $\langle\tau^z\rangle_G\simeq
0.32$. The orbital polarization in LaVO$_3$ 
$\langle\tau^x\rangle\simeq 0.03$ is rather weak at $T_{N1}$,
and is further reduced with decreasing $T<T_{\rm OO}$. Note that also
spin order parameter is expected to be reduced below 
$\langle S^z\rangle=1$, but weak quantum fluctuations in the $C$-AF 
phase \cite{Rac02} were neglected here. In contrast, 
in SmVO$_3$ the phase transitions separate: the orbital transition
occurs first at $T_{\rm OO}\simeq 0.86J$, and the magnetic one follows 
at a lower $T_{N1}\simeq 0.65J$. Already in this case the transverse 
orbital polarization is considerably increased, with
$\langle\tau^x\rangle\simeq 0.20$ at $T_{N1}$ (see Fig. \ref{fig:ops}),
and further increases with decreasing $r_R$ (not shown). Note that the 
polarization $\langle\tau^x\rangle$ does not change close to
$T_{\rm OO}$, and only below $T_{N1}$ gets weakly reduced due to the 
developing magnetic order, in agreement with experiment \cite{Sag07}.
The $G$-OO parameter is here stronger as the singlet orbital 
fluctuations are not so pronounced when $T\to 0$,
being $\langle\tau^z\rangle_G\simeq 0.37$.  

\begin{figure}[t!]
\begin{center}
\begin{minipage}{0.49\textwidth}
    \includegraphics[width=\textwidth]{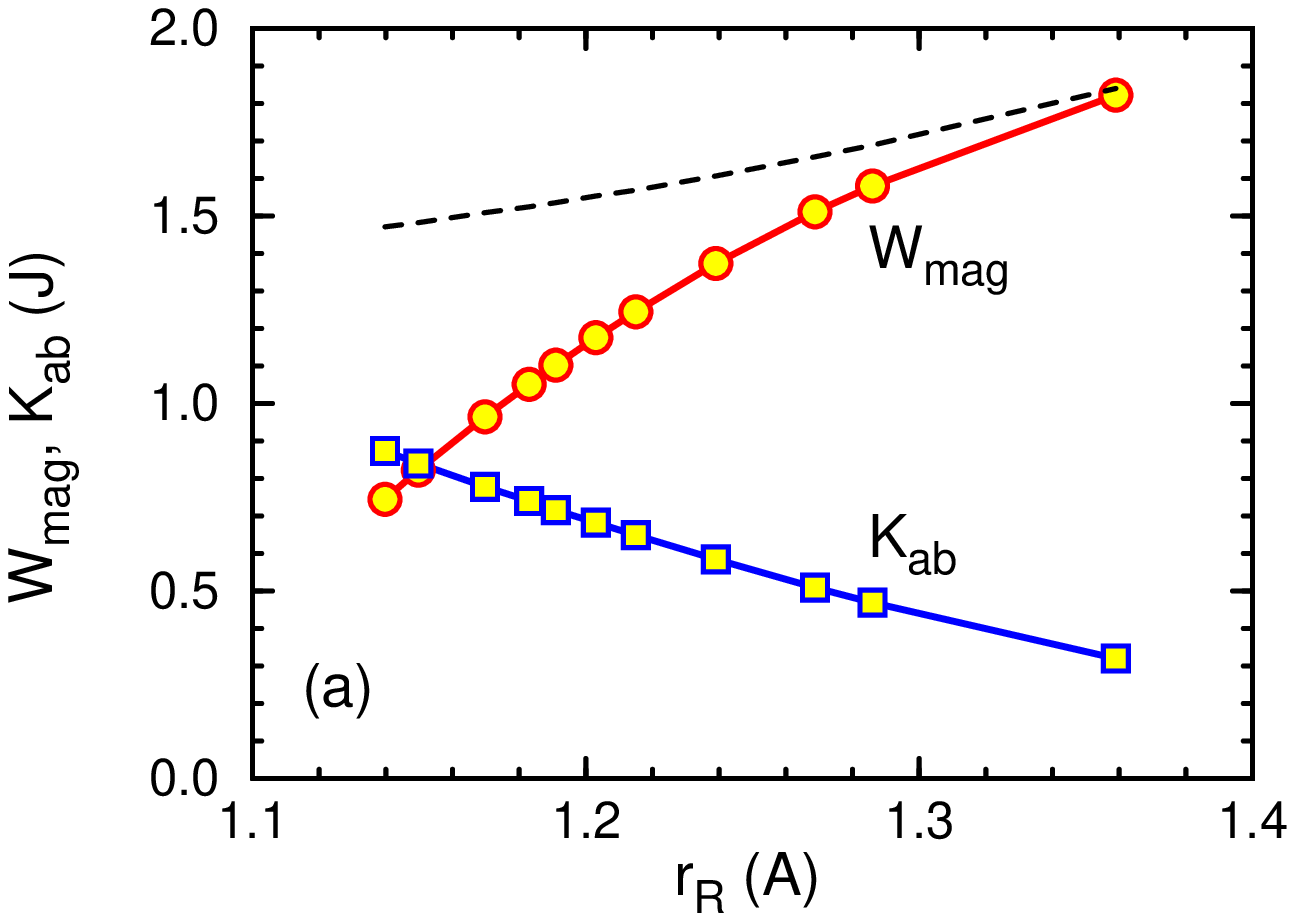}
\end{minipage}
 \quad
\begin{minipage}{0.43\textwidth}
    \includegraphics[width=\textwidth]{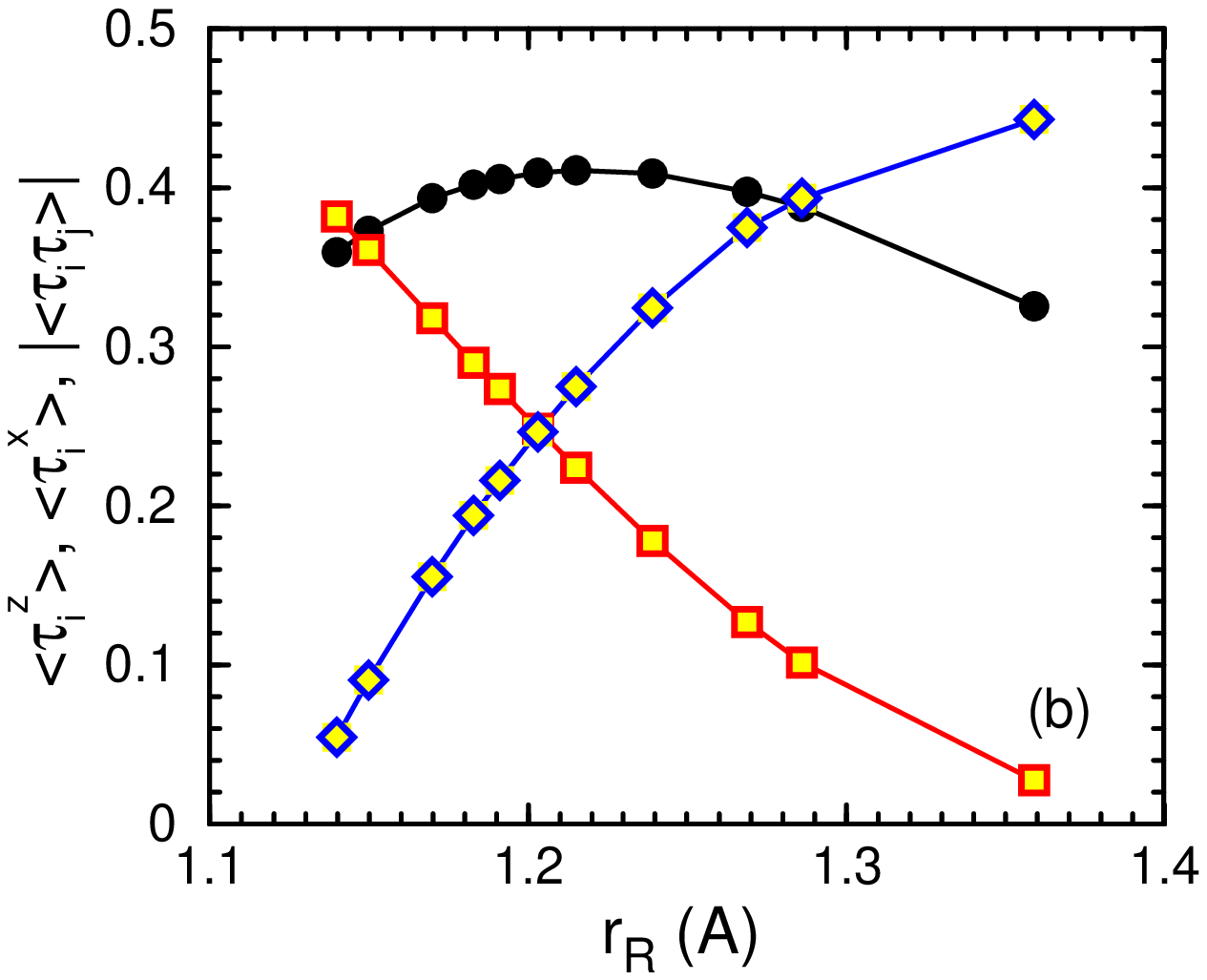}
\end{minipage}
\end{center}
\caption{
(a) The width of magnon band $W_{C-{\rm AF}}$ for finite $g_{\rm eff}$
(circles) and without orbital-strain coupling ($g_{\rm eff}=0$,
dashed), and orbital interactions in $ab$ planes $K_{ab}$ (squares)
in the $C$-AF phase of cubic vanadates (the points correspond to
the $R$VO$_3$ compounds of Fig. \ref{fig:phd}).
(b) Evolution of the orbital order parameter $\langle\tau_i^z\rangle_G$
(filled circles), transverse orbital polarization 
$\langle\tau_i^x\rangle$ (squares), and orbital intersite correlations
$|\langle\vec{\tau}_i\cdot\vec{\tau}_j\rangle|$ (diamonds) along
$c$ axis at $T=0$.
Parameters: $v_z=17$, $v_{ab}=22$, $v_{g}=740$.}
\label{fig:orbi}
\end{figure}

The key features of the present spin-orbital system which drive the
observed dependence of $T_{\rm OO}$ and $T_{N1}$ on $r_R$ \cite{Miy03}
is the evolution of intersite orbital correlations are: 
($i$) the gradual increase of the orbital interactions  
$K_{ab}\tau^z_i\tau^z_j$ [Fig. \ref{fig:orbi}(a)], and 
($ii$) the reduction of orbital fluctuations on the bonds along the 
$c$ axis, described by the bond singlet correlations
$\langle{\vec\tau}_i\cdot{\vec\tau}_j\rangle$ [Fig. \ref{fig:orbi}(b)].
The parameter $K_{ab}$ in Fig. \ref{fig:orbi}(a) consists of the 
superexchange contribution $\propto J$ (\ref{orbka}) and orbital 
interaction $V_{ab}$ (\ref{HJT}) induced by the lattice distortion. 
While the superexchange does not change with 
decreasing $r_R$, the latter term increases and induces the increase
of $T_{\rm OO}$ from LaVO$_3$ to YVO$_3$. This increase is similar to 
that observed in the $R$MnO$_3$ manganites \cite{Goo06}. Thereby the 
bond angle $\Theta$ decreases from $157.4^{\circ}$ in LaVO$_3$ to 
$144.8^{\circ}$ in YVO$_3$.

While the singlet correlations are drastically suppressed from LaVO$_3$ 
towards LuVO$_3$, the orbital order parameter
$\langle\tau^z\rangle_G$ somewhat increases from LaVO$_3$ to SmVO$_3$
(see also Fig. \ref{fig:ops}). At the same time 
the orbital polarization $\langle\tau^x\rangle$ increases and soon 
becomes as important as the orbital order parameter, i.e., $\langle\tau^x\rangle\simeq\langle\tau^z\rangle_G$. Further increase of 
the orbital polarization towards LuVO$_3$ suppresses the $G$-AO 
parameter, so $\langle\tau^z\rangle_G$ passes through a maximum and
decreases for $r_R<1.22$ \AA{}. 

It is remarkable that the above changes in orbital correlations induced 
by the lattice suppress gradually the magnetic interactions in the 
$C$-AF phase, although the value of $J$ remains unchanged. This is well
visible in the total width of the magnon band, 
$W_{C-{\rm AF}}=4(J_{ab}+|J_c|)$ (at $T=0$) \cite{Ole07}, shown in Fig.
\ref{fig:orbi}(a), being reduced from $\sim 1.84J$ in LaVO$_3$ to 
$\sim 1.05J$ in YVO$_3$. This large reduction qualitatively agrees with
the rather small values of the exchange constants in the $C$-AF
phase of YVO$_3$ \cite{Ulr03}, see also Fig. \ref{fig:swcafd}. This 
reduction is caused by the suppression of the singlet orbital 
correlations $\langle{\vec\tau}_i\cdot{\vec\tau}_j\rangle$ by the 
increasing coupling to the lattice $g_{\rm eff}(\vartheta)$ when 
$r_R$ decreases. Note also that this effect would be rather small for
$g_{\rm eff}=0$ --- this behavior is excluded by experiment.

\begin{figure}[t!]
\sidecaption[t]
\includegraphics[width=6.8cm]{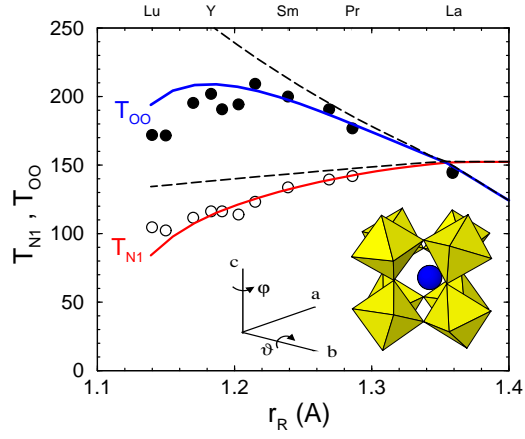}
\caption{
The orbital $T_{\rm OO}$ and magnetic $T_{N1}$ transition temperature
for varying $r_R$ in the $R$VO$_3$ perovskites, obtained from model
(\ref{rvo3}) for: $v_{g}=740$ (solid lines) and $v_g=0$ (dashed lines). 
Circles show the experimentat data of Ref. \cite{Miy03}. 
The inset shows the GdFeO$_3$-type distortion, with the rotation angles
$\vartheta$ and $\varphi$. Other parameters as in Fig. \ref{fig:orbi}.
This figure is reproduced from Ref. \cite{Hor08}.
}
\label{fig:phd}
\end{figure}

Following Ref. \cite{Hor08}, we argue that the gradual reduction of the 
orbital singlet correlations in favor of increasing orbital polarization 
is responsible for the evolution of the orbital transition temperature
$T_{\rm OO}$ in the experimental phase diagram of Fig. \ref{fig:phd},
which is reproduced by the theory in the entire range of available 
$r_R$. The transition temperature $T_{\rm OO}$ changes in a nonmonotonic 
way, similar to the orbital order parameter $\langle\tau^z\rangle_G$ at
$T=0$ [Fig. \ref{fig:orbi}(b)]. After analyzing the changes in the 
orbital correlations, we see that the physical reasons of the decrease
of $T_{\rm OO}$ for small (large) $r_R$ are quite different. While the 
orbital fluctuations dominate and largely suppress the orbital order in 
LaVO$_3$, the orbital polarization takes over near YVO$_3$ and competes
with $G$-AO order.

\begin{figure}[b!]
\sidecaption[t]
\includegraphics[width=7.5cm]{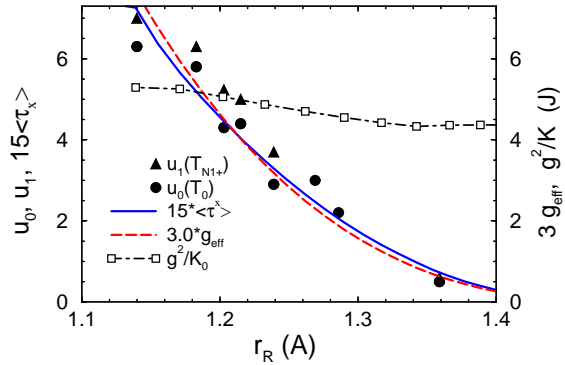}
\caption{
Experimental distortion (in percent) at $T=0$ ($u_0$, circles)
and above $T_{N1}$ ($u_1$, triangles) for the
$R$VO$_3$ compounds \cite{Ren03,Ree06,Sag07},
compared with the orbital polarization
$\langle\tau^x\rangle_{T=0}$ and with $g_{\rm eff}$ (\ref{geff});
$g_{\rm eff}$ and $g^2/K$ are in units of $J$.
Squares show the upper bound for $g^2/K$ predicted by the theory.
Parameters: $v_z=17$, $v_{ab}=22$, $v_{g}=740$.
This figure is reproduced from Ref. \cite{Hor08}.
}
\label{fig:u}
\end{figure}

While the above fast dependence on the tilting angle $\vartheta$ of
VO$_6$ octahedra in the $R$VO$_3$ family was introduced in order to
reproduce the experimentally observed dependence of $T_{\rm OO}$ on
$r_R$, see Fig. \ref{fig:phd}, it may
be justified {\it a posteriori\/}. It turns out that
the dependence of $g_{\rm eff}$ on the ionic radius $r_R$ in Eq.
(\ref{geff}) follows the actual lattice distortion $u$ in $R$VO$_3$
measured at $T=0$ ($u_0$), or just above $T_{N1}$ ($u_1$) \cite{Hor08}. 
Also the orbital polarization $\langle\tau^x\rangle$ is approximately
$\propto\sin^5\vartheta\cos\vartheta$, and follows the same fast 
dependence of $g_{\rm eff}(\vartheta$ for the $R$VO$_3$ perovskites
(Fig. \ref{fig:u}).
This result is somewhat unexpected, as information about the actual 
lattice distortions has not been used in constructing the miscroscopic 
model (\ref{som}).
These results indicate that the bare coupling parameters $\{g,K\}$ are
nearly constant and independent of $r_R$, which may be treated as a
prediction of the theory to be verified by future experiments.

\section{Summary and outlook}
\label{sec:summa}

\begin{svgraybox}
Summarizing, spin-orbital superexchange model (\ref{rvo3}) augmented
by orbital-lattice couplings provides an explanation of the 
experimental variation of the orbital $T_{\rm OO}$ and magnetic 
$T_{N1}$ transition temperatures for the whole class of the $R$VO$_3$ 
perovskites. A more complete theoretical understanding including 
a description of the second magnetic transition from 
$C$-AF to $G$-AF phase, which occurs at $T_{N2}$ for small ionic 
radii $r_R$ \cite{Maz08}, remains to be addressed by future theory, 
which should include the spin-orbit relativistic coupling \cite{Hor03}.
\end{svgraybox}

We conclude by mentioning a few open issues and future
directions of reasearch in the field of perovskite vanadates.
Rapid progress of the field of orbital physics results mainly from 
experiment, and is triggered by the synthesis of novel materials. 
Although experiment is ahead of theory in most cases, there are some
exceptions. One of them was a theoretical prediction of the energy 
and dispersion of orbital excitations \cite{vdB99,Ish97,vdB01}.
Only recently orbital excitations (orbitons) could be observed by
Raman scattering in the Mott insulators LaTiO$_3$ and YTiO$_3$ 
\cite{Ulr06,Ulr08}. They were also identified in the optical absorption
spectra of YVO$_3$ and HoVO$_3$ \cite{Ben08}. The exchange of two 
orbitals along the $c$ axis in the intermediate $C$-AF phase was shown 
to contribute to the optical conductivity $\sigma(\omega)$.

An interesting question which arises in this context is the carrier 
propagation in a Mott insulator with orbital order. This problem is 
rather complex as in a spin-orbital polaron, created by doping of a
single hole, both spin and orbital excitations contribute to the hole
scattering \cite{Zaa93}, which may even become localized by string 
excitations as in the $t$-$J^z$ model \cite{Mar91}.
Indeed, the coupling to orbitons increases 
the effective mass of a moving hole in $e_g$ systems \cite{vdB00}. 
The orbital part of the superexchange is classical (compass-like) 
in $t_{2g}$ systems, but nevertheless the hole is not confined as weak 
quasiparticle dispersion arises from three-site processes 
\cite{Dag08,Woh08}. 

As in the doped manganites, also in doped $R_{1-x}$(Sr,Ca)$_x$VO$_3$ 
systems the $G$-AO order gradually disappears \cite{Fuj05}. The 
$C$-AF spin order survives, however, in a broad range of doping, in 
contrast to La$_{1-x}$Sr$_x$MnO$_3$, where FM order replaces the $A$-AF
phase already at $x\sim 0.10$, and is accompanied by the 
$e_g$ orbital liquid \cite{Ole02} at higher doping. It is quite 
remarkable that the complementary $G$-AF/$C$-AO order is fragile and
disappears in Y$_{1-x}$Ca$_x$VO$_3$ already at $x=0.02$ \cite{Fuj05}. 
The doped holes in $C$-AF/$G$-AO phase are localized in polaron-like 
states \cite{Fuj08}, so the pure electronic model such as the one of 
Ref. \cite{Dag08} is too crude to capture both the evolution of the 
spin-orbital order in doped vanadates and the gradual decrease of the 
energy scale for spin-orbital fluctuations. Theretical studies at finite
hole concentration are still nonexistent in 3D models, but one may 
expect a transition from a phase with AF order to a phase with FM spin  
polarization at large Hund's coupling, as shown both for $e_g$
\cite{Dag04} and $t_{2g}$ \cite{Fuj05} systems.

A few representative problems related to the properties of $R$VO$_3$ 
perovskites discussed above demonstrate that the
orbital physics is a very rich field, with intrinsically frustrated
interactions and rather exotic ordered or disordered phases, with their 
behavior dominated by quantum fluctuations. While valuable information 
about the electronic structure is obtained from density functional 
theory \cite{Sol08}, the many-body aspects have to be studied 
simultaneously using models of correlated electrons. The $R$VO$_3$ 
perovskites remain an interesting field of research, as it turned out 
that electron-lattice coupling is here not strong enough to suppress 
(quench) the orbital fluctuations \cite{Hor08}. 
Thus the composite quantum fluctuations described by the spin-orbital 
model (\ref{rvo3}) remain active. Nevertheless, there is significant 
control of the electronic properties due to the electron-lattice 
coupling. Thus, the lattice distortions may also influence the 
onset of magnetic order in systems with active orbital degrees of 
freedom. If they are absent and the lattice is frustrated in addition, 
a very interesting situation arises, with strong tendency towards truly 
exotic quantum states \cite{Kha05}. Examples of this behavior were 
considered recently for the triangular lattice,
both for $e_g$ orbitals in LiNiO$_2$ \cite{Rei05} and $t_{2g}$ orbitals
in NaTiO$_2$ \cite{Nor08}. None of these models could really be solved,
but generic tendency towards dimer correlations with spin singlets on
the bonds for particular orbital states has been shown. Yet, the
question whether novel types of orbital order, such as e.g. nematic 
order in spin models \cite{Sha06}, could be found in certain situations 
remains open.

\begin{acknowledgement}
It is our great pleasure to thank G. Khaliullin and
L.F. Feiner for very stimulating collaboration which
significantly contributed to our present understanding of the subject.
We thank B. Keimer, G.A. Sawatzky, Y. Tokura and particularly C. Ulrich
for numerous insightful discussions.
A.M. Ole\'s acknowledges financial support by
the Foundation for Polish Science (FNP) and by the Polish
Ministry of Science and Education under Project No. N202 068 32/1481.

\end{acknowledgement}


\begin{thebibliography}{99}


\bibitem{Tok00} Y. Tokura, N. Nagaosa,
                   Science \textbf{288}, 462 (2000)

\bibitem{Fei99} L.F. Feiner, A.M. Ole\'s,
                   Phys. Rev. B \textbf{59}, 3295 (1999)

\bibitem{Kha01} G. Khaliullin, P. Horsch, A.M. Ole\'s,
                   Phys. Rev. Lett. \textbf{86}, 3879 (2001)

\bibitem{Ole05} A.M. Ole\'s, P.~Horsch, G. Khaliullin, L.F. Feiner,
                   Phys. Rev. B \textbf{72}, 214431 (2005)

\bibitem{Miy05} S. Miyasaka, S. Onoda, Y. Okimoto, J. Fujioka, M. Iwama,
                   N. Nagaosa, Y. Tokura,
                   Phys. Rev. Lett. \textbf{94}, 076405 (2005)

\bibitem{Yan07} J.-Q. Yan, J.-S. Zhou, J.B. Goodenough, Y. Ren, 
                   J.G. Cheng, S. Chang, J. Zarestky, O. Garlea, 
                   A. Liobet, H.D. Zhou, Y. Sui, W.H. Su, R.J. McQueeney,
                   Phys. Rev. Lett. \textbf{99}, 197201 (2007)

\bibitem{Kug82} K.I. Kugel, D.I. Khomskii,
                   Sov. Phys. Usp. \textbf{25}, 231 (1982)

\bibitem{Fei97} L.F. Feiner, A.M. Ole\'s, J. Zaanen,
                   Phys. Rev. Lett. \textbf{78}, 2799 (1997)

\bibitem{vdB04} J. van den Brink,
                   New J. Phys. \textbf{6}, 201 (2004)

\bibitem{Ole06} A.M. Ole\'s, P.~Horsch, L.F. Feiner, G. Khaliullin,
                   Phys. Rev. Lett. \textbf{96}, 147205 (2006)

\bibitem{Fri99} B. Frishmuth, F. Mila, M. Troyer,
                   Phys. Rev. Lett. \textbf{82}, 835 (1999)

\bibitem{vdB99} J. van der Brink, P. Horsch, F. Mack, A.M. Ole\'s,
                   Phys. Rev. B \textbf{59}, 6795 (1999)

\bibitem{Dag04} M. Daghofer, A.M. Ole\'s, W. von der Linden,
                   Phys. Rev. B \textbf{70}, 184430 (2004)

\bibitem{Fei05} L.F. Feiner, A.M. Ole\'s,
                   Phys. Rev. B \textbf{71}, 144422 (2005)

\bibitem{Kha05} G. Khaliullin,
                   Prog. Theor. Phys. Suppl. \textbf{160}, 155 (2005)

\bibitem{Kho03} D.I. Khomskii, M.V. Mostovoy,
                   J. Phys. A \textbf{36}, 9197 (2003)

\bibitem{Mil05} J. Dorier, F. Becca, F. Mila,
                   Phys. Rev. B  \textbf{72}, 024448 (2005)

\bibitem{Brz07} W. Brzezicki, J. Dziarmaga, A.M. Ole\'s,
                   Phys. Rev. B  \textbf{75} 134415 (2007)

\bibitem{Ulr03} C. Ulrich, G. Khaliullin, J. Sirker, M. Reehuis, M. Ohl, 
                   S. Miyasaka, Y. Tokura, B. Keimer,
                   Phys. Rev. Lett. \textbf{91}, 257202 (2003)

\bibitem{Miy03} S. Miyasaka, Y. Okimoto, M. Iwama, Y. Tokura,
                   Phys. Rev. B \textbf{68}, 100406 (2003)

\bibitem{Goo06} J.-S. Zhou, J.B. Goodenough,
                   Phys. Rev. Lett. \textbf{96}, 247202 (2006)

\bibitem{Miy06} S. Miyasaka, J. Fujioka, M. Iwama, Y. Okimoto, 
                   Y. Tokura,
                   Phys. Rev. B \textbf{73}, 224436 (2006)

\bibitem{Fei98} L.F. Feiner, A.M. Ole\'s, J. Zaanen,
                   J. Phys.: Condens. Matter {\bf 10}, L555 (1998)

\bibitem{Zaa93} J. Zaanen, A.M. Ole\'s,
                   Phys. Rev. B \textbf{48}, 7197 (1993)

\bibitem{Ole07} A.M. Ole\'s, P. Horsch, G. Khaliullin,
                   Phys. Rev. B \textbf{75}, 184434 (2007)

\bibitem{Gri71} J.S. Griffith,
                   {\it The Theory of Transition Metal Ions\/}
                   (Cambridge University Press, Cambridge, 1971)

\bibitem{Wei04} A Wei\ss{}e, H. Fehske,
                   New J. Phys. {\bf 6} 158 (2004)

\bibitem{Ole00} A.M. Ole\'s, L.F. Feiner, J. Zaanen,
                   Phys. Rev. B \textbf{61}, 6257 (2000)

\bibitem{And07} M. De Raychaudhury, E. Pavarini, O.K. Andersen,
                   Phys. Rev. Lett. \textbf{99}, 126402 (2007)

\bibitem{Hor08} P. Horsch, A.M. Ole\'s, L.F. Feiner, G. Khaliullin,
                   Phys. Rev. Lett. \textbf{100}, 167205 (2008)

\bibitem{Hor03} P. Horsch, G. Khaliullin, A.M. Ole\'s,
                   Phys. Rev. Lett. \textbf{91}, 257203 (2003)

\bibitem{Kha04} G. Khaliullin, P. Horsch, A.M. Ole\'s,
                   Phys. Rev. B \textbf{70}, 195103 (2004)

\bibitem{Bae86} D. Baeriswyl, J. Carmelo, A. Luther,
                   Phys. Rev. B \textbf{33}, 7247 (1986)

\bibitem{Aic02} M. Aichhorn, P. Horsch, W. von der Linden, M. Cuoco, 
                   Phys. Rev. B \textbf{65}, 201102 (2002)

\bibitem{Miy02} S. Miyasaka, Y. Okimoto, Y. Tokura,
                   J. Phys. Soc. Jpn. \textbf{71}, 2086 (2002)

\bibitem{Ren00} Y. Ren, T.T.M. Palstra, D.I. Khomskii, A.A. Nugroho,
                   A.A. Menovsky, G.A. Sawatzky,
                   Phys. Rev. B \textbf{62}, 6577 (2000)

\bibitem{Sir03} J. Sirker, G. Khaliullin,
                   Phys. Rev. B \textbf{67}, 100408(R) (2003)

\bibitem{Sir08} J. Sirker, A. Herzog, A.M. Ole\'s, P. Horsch,
                   Phys. Rev. Lett. \textbf{101}, 157204 (2008)

\bibitem{Rac02} M. Raczkowski, A.M. Ole\'s, 
                   Phys. Rev. B \textbf{66}, 094431 (2002)

\bibitem{Ren03} Y. Ren, A.A. Nugroho, A.A. Menovsky, J. Strempfer, 
                   U. R\"utt, F. Iga, T. Takabatake, C.W. Kimball,
                   Phys. Rev. B \textbf{67}, 014107 (2003)

\bibitem{Ree06} M. Reehuis, C. Ulrich, P. Pattison, B. Ouladdiaf, 
                   M.C. Rheinst\"adter, M. Ohl, L.P. Regnault, 
                   M. Miyasaka, Y. Tokura, B. Keimer,
                   Phys. Rev. B \textbf{73}, 094440 (2006)

\bibitem{Sag06} M.H. Sage, G.R. Blake, G.J. Nieuwenhuys, T.T.M. Palstra,
                   Phys. Rev. Lett. \textbf{96}, 036401 (2006)

\bibitem{Sag07} M.H. Sage, G.R. Blake, C. Marquina, T.T.M. Palstra,
                   Phys. Rev. B \textbf{76}, 195102 (2007)

\bibitem{Pav05} E. Pavarini, A. Yamasaki, J. Nuss, O.K. Andersen,
                   New J. Phys. \textbf{7}, 188 (2005)

\bibitem{Dit80} R.V. Ditzian, J.R. Banavar, G.S. Grest, L.P. Kadanoff,
                   Phys. Rev. B \textbf{22}, 2542 (1980)

\bibitem{Silva} T.N. De Silva, A. Joshi, M. Ma, F.C. Zhang,
                   Phys. Rev. B \textbf{68}, 184402 (2003)

\bibitem{Maz08} D.A. Mazurenko, A.A. Nugroho, T.T.M. Palstra, 
                   P.H.M. van Loosdrecht,
                   Phys. Rev. Lett. \textbf{101}, 245702 (2008)
 
\bibitem{Ish97} S. Ishihara, J. Inoue, S. Maekawa,
                   Phys. Rev. B \textbf{55}, 8280 (1997)

\bibitem{vdB01} J. van~den Brink,
                   Phys. Rev. Lett. \textbf{87}, 217202 (2001)
 
\bibitem{Ulr06} C. Ulrich, A. G\"ossling, M. Gr\"uninger, M. Guennou, 
                   H. Roth, M. Cwik, T. Lorenz, G. Khaliullin,
                   B. Keimer,
                   Phys. Rev. Lett. \textbf{97}, 157401 (2006)

\bibitem{Ulr08} C. Ulrich, G. Ghiringhelli, A. Piazzalunga, L. Braicovich, 
                   N.B. Brookes, H. Roth, T. Lorenz, B. Keimer,
                   Phys. Rev. B \textbf{77}, 113102 (2008)

\bibitem{Ben08} E. Benckiser, R. R\"uckamp, T. M\"oller, T. Taetz, 
                   A. M\"oller, A.A. Nugroho, T.T.M. Palstra, 
                   G.S. Uhrig, M. Gr\"uninger,
                   New J. Phys. \textbf{10}, 053027 (2008)

\bibitem{Mar91} G. Mart\'inez, P. Horsch,
                   Phys. Rev. B \textbf{44}, 317 (1991)

\bibitem{vdB00} J. van~den Brink, P. Horsch, A.M. Ole\'s,
                   Phys. Rev. Lett. \textbf{85}, 5174 (2000)

\bibitem{Dag08} M. Daghofer, K. Wohlfeld, A.M. Ole\'s, E. Arrigoni,
                   P. Horsch,
                   Phys. Rev. Lett. \textbf{100}, 066403 (2008)

\bibitem{Woh08} K. Wohlfeld, M. Daghofer, A.M. Ole\'s, P. Horsch,
                   Phys. Rev. B \textbf{78}, 214423 (2008)

\bibitem{Fuj05} J. Fujioka, S. Miyasaka, Y. Tokura,
                   Phys. Rev. B \textbf{72}, 024460 (2005)

\bibitem{Ole02} A.M. Ole\'s, L.F. Feiner, 
                   Phys. Rev. B \textbf{65}, 052414 (2000)

\bibitem{Fuj08} J. Fujioka, S. Miyasaka, Y. Tokura,
                   Phys. Rev. B \textbf{77}, 144402 (2008)

\bibitem{Dam08} J. Sirker, J. Damerau, A. Kl\"umper,
                   Phys. Rev. B \textbf{78}, 235125 (2008)

\bibitem{Sol08} I.V. Solovyev,
                   J. Phys.: Condens. Matter \textbf{20}, 293201 (2008). 

\bibitem{Rei05} A.J.W. Reitsma, L.F. Feiner, A.M. Ole\'s,
                   New J. Phys. \textbf{7}, 121 (2005)

\bibitem{Nor08} B. Normand, A.M. Ole\'s,
                   Phys. Rev. B \textbf{78}, 094427 (2008)

\bibitem{Sha06} N. Shannon, T. Momoi, P. Sindzingre,
                   Phys. Rev. Lett. \textbf{96}, 027213 (2006)


\end{thebibliography}
\end{document}